\documentclass[preprint,12pt]{article}
 
\usepackage{amssymb}
\usepackage{caption}
\usepackage{breqn}
\usepackage{graphicx}
\usepackage{breqn}

\begin{document}
 
 \renewcommand{\theequation}{\arabic{equation}}    
  % redefine the command that creates the equation no.    
  \setcounter{equation}{0}  % reset counter   
    \setcounter{section}{0}  % reset counter 

\textbf{Reflection Identities of Harmonic Sums and pole decomposition of BFKL eigenvalue}
\begin{center}

\small
Mohammad Joubat$^{(a)}$   and Alex Prygarin$^{(b)}$  
\\
$^{(a)}$ Department of Mathematics, Ariel University, Ariel 40700, Israel\\
$^{(b)}$ Department of Physics, Ariel University, Ariel 40700, Israel 
\end{center}
\normalsize

\normalsize

\begin{abstract}
   We analyze  known results of next-to-next-to-leading~(NNLO) singlet BFKL eigenvalue in $N=4$ SYM written in terms of harmonic sums. The nested harmonic sums building   known NNLO BFKL eigenvalue for specific values of the conformal spin have poles at negative integers. We sort the harmonic sums according to the complexity with respect to their weight and depth and use their pole decomposition in terms of the reflection identities to find the most complicated terms of NNLO BFKL eigenvalue for an arbitrary value of the conformal spin. The obtained result is compatible with the Bethe-Salpeter approach to the BFKL evolution.

\end{abstract}
%%%%%%%%%%%%%%%%%%%%%%%%%%%%%%%%%%%%%%%%%%%%%%%%%%%%%%%%%%%
%%%%%%%%%%%%%%%%%%%%%%%%%%%%%%%%%%%%%%%%%%%%%%%%%%%%%%%%%%%
%%%%%%%%%%%%%%%%%%%%%%%%%%%%%%%%%%%%%%%%%%%%%%%%%%%%%%%%%%%
%%%%%%%%%%%%%%%%%%%%%%%%%%%%%%%%%%%%%%%%%%%%%%%%%%%%%%%%%%%
%%%%%%%%%%%%%%%%%%%%%%%%%%%%%%%%%%%%%%%%%%%%%%%%%%%%%%%%%%%
%%%%%%%%%%%%%%%%%%%%%%%%%%%%%%%%%%%%%%%%%%%%%%%%%%%%%%%%%%%
%%%%%%%%%%%%%%%%%%%%%%%%%%%%%%%%%%%%%%%%%%%%%%%%%%%%%%%%%%%
%%%%%%%%%%%%%%%%%%%%%%%%%%%%%%%%%%%%%%%%%%%%%%%%%%%%%%%%%%%
%%%%%%%%%%%%%%%%%%%%%%%%%%%%%%%%%%%%%%%%%%%%%%%%%%%%%%%%%%%

\section{Introduction}\label{intro}

In the present study we continue to discuss the eigenvalue of the Balitsky-Fadin-Kuraev-Lipatov~(BFKL) equation~\cite{BFKL} in $N=4$ super Yang-Mills~(SYM) theory. 
Despite the fact that the maximally supersymmetric $N=4$ SYM is a more complicated theory than Quantum Chromodynamics~(QCD), the calculations and  results   in $N=4$ SYM are often much simpler than  that  of QCD.  The traditional techniques of loop calculations using Feynman diagrams are of rapidly increasing complexity with each loop order suggesting new and more advanced techniques to be developed for the perturbative calculations. Recent developments in the integrability approach~(see a review by  Alfimov, Gromov and  Kazakov~\cite{Alfimov:2020obh})  allowed to  calculate the next-to-next-to-leading~(NLO) order color singlet  BFKL eigenvalue in $N=4$ SYM   in different regions of parameters~(conformal spin and anomalous dimension). However, the full closed functional form of the NNLO BFKL eigenvalue is still to be found. The leading order~(LO) and the next-to-leading~(NLO) order BFKL eigenvalue  in QCD in the arbitrary color configuration   is known for many years~\cite{NLO} . In $N=4$ SYM the leading order BFKL eigenvalue coincides~(up to coupling constant redefinition) with the  QCD expression, while the NLO eigenvalue in $N=4$ SYM is built of the most complicated functions of the corresponding QCD result. The last fact is known as the property of \emph{maximal transcendentality} formulated by Kotikov and Lipatov~\cite{Kotikov:2001sc,Kotikov:2002abmaxtrans} that  allows to significantly simplify calculations as we show in the present paper.

The main goal of our study is to analyze the most complicated part of the NNLO BFKL eigenvalue based on the available results and to derive its explicit functional form for  arbitrary values of the  conformal spin and the anomalous dimension.  We use the pole decomposition of the cross products of harmonic sums of  conjugate arguments and the reflection identities to reconstruct the full functional form of the most complicated part of the color singlet  NNLO BFKL eigenvalue in $N=4$ SYM.  

The paper is organized as follows. In the first section we discuss the of state of art of the perturbative  calculations of the BFKL eigenvalue and define the harmonic sums. In the next section we show how the color singlet NLO BFKL eigenvalue for specific values of conformal spin ($n=0$ and $n=1$)  can be decomposed using the reflection identities of harmonic sums into separate pieces having poles on either negative or non-negative values of the complex argument $z=-\frac{1}{2}+i \frac{\nu}{2}+\frac{n}{2}$. We explain the importance of this decomposition and compare our results to  known expressions.  The Conclusions and Discussions section summarizes the main results. The relevant reflection identities as well as some other technical details of our calculations are presented in Appendix.

\section{BFKL eigenvalue and  harmonic sums}
 In this section we follow the lines of works of   Gromov, Levkovich-Maslyuk and Sizov~\cite{GROMOV}, Caron-Huot and Herranen~\cite{HUOT}, and Alfimov, Gromov and Sizov \cite{GromovNonzero}
 and discuss how that the known leading order~(LO) and the next-to-leading order~(NLO)   BFKL eigenvalues in $N=4$ SYM can be expressed in terms of the nested harmonic sums analytically continued from positive integer values of the argument to the complex plane. 
 % The analytic continuation is performed according to the prescription formulated by Kotikov and Velizhanin~\cite{Velizhanin}. 
 
 The nested harmonic sums are defined~\cite{HS1,Vermaseren:1998uu,Blumlein:1998if,Remiddi:1999ew} as  nested summation for $n\in \mathbb{N}$
\begin{eqnarray}\label{defS}
S_{a_1,a_2,...,a_k}(n)=  \sum_{n \geq i_1 \geq i_2 \geq ... \geq i_k \geq 1 }   \frac{\mathtt{sign}(a_1)^{i_1}}{i_1^{|a_1|}}... \frac{\mathtt{sign}(a_k)^{i_k}}{i_k^{|a_k|}}
\end{eqnarray}

We are interested only  in  harmonic sums with real integer values of $a_i$, which build the alphabet of the possible negative and positive indices and   uniquely  label them.  
For a given harmonic sum $S_{a_1,a_2,...,a_k}(n)$ the index $k$ denotes  the \emph{depth} while  $w=\sum_{i=1}^{k}|a_i|$ is called the  \emph{weight} of the harmonic sum  

The harmonic sums $S_{a_1, a_2, ..., a_n} (z)$ can be analytically continued from   integers values of the argument to the complex plane as discussed by Kotikov and Velizhanin~\cite{Velizhanin}.  There are two different analytic continuations of
the harmonic sums  to the complex plane  $a)$ the analytic
continuation from the even integer values of the argument and $b)$ the analytic continuation from the odd integer values.  We follow the authors  of Ref.~\cite{GROMOV} and  use the analytic continuation of the harmonic sums from the even integer values of the argument, which is implemented in their dedicated Mathematica package. After the analytic continuation all the corresponding  meromorphic functions  \emph{the same pole structure}, namely,  poles at the negative integer values of the argument and the maximal pole order equals the weight of the  corresponding harmonic sum.

The BFKL eigenvalue is the function of two transverse degrees of freedom that originates from the longitudinal and transverse momenta separation in the multi-Regge kinematics. The BFKL eigenvalue explicitly depends on the complex variable 
\begin{eqnarray}
z=-\frac{1}{2}+\frac{i \nu}{2}+\frac{n}{2},
\end{eqnarray} 
that is obtained from the two dimensional  Mellin transform of the BFKL amplitude. 
Here $\nu$ is the real continuous variable that is related to  the anomalous dimension of twist-2 operators, while $n$ takes values of real integers and has a meaning of  conformal spin.  The harmonic sums $S_{\alpha}(z)$ have poles at  imaginary integer  values of $\nu$, namely
\begin{eqnarray}\label{nu}
\nu =i (2 k +n -1)
\end{eqnarray}
for $k=1,2,3,..$. 
It should be emphasized that all harmonic sums of the complex conjugate variable $\bar{z}$ are  finite at the values of $\nu$ given in eq.~(\ref{nu}). This rather simple statement allows us to develop a powerful technique of pole decomposition based  the fact that only a limited set of the transcendental constants can emerge at any given weight. It should be mentioned that our method only a slight modification of the one implemented by  Gromov, Levkovich-Maslyuk and Sizov~\cite{GROMOV} in their Mathematica package in calculating  the next-to-next-to-leading~(NNLO) order singlet BFKL eigenvalue in $N=4$ SYM.      Our modification comes the explicit calculation of the reflection identities up to weight five, which effectively decompose the product of two harmonic sums $s_{\{\alpha\}}(z)s_{\{\beta\}}(\bar{z})$ with mixed pole structure into a sum of terms that have separately poles at either right or left semiplanes of the complex $z$ plane. This can be  schematically depicted as follows 
\begin{eqnarray}\label{shuffle}
s_{\{\alpha\}}(z)s_{\{\beta\}}(\bar{z})=\sum_{\{\gamma\}} a_{\{\gamma\}} s_{\{\gamma\}}(z)\;\; +\;\; 
\sum_{\{\lambda\}} b_{\{\gamma\}} s_{\{\lambda\}}(\bar{z}),  
\end{eqnarray}
where
the summation is performed over all possible letter configurations of harmonic sums at given weight including the harmonic sums at lower weights times the corresponding transcendental constants, and the coefficients $a_{\{\gamma\}}$ and $b_{\{\gamma\}}$ are rational numbers.

The BFKL eigenvalue  $\omega(z,\bar{z}) $ in $N=4$ SYM in the singlet color configuration can be written as follows 
\begin{eqnarray}\label{gnexp}
\omega(z,\bar{z})& = &  \omega_{0}(z,\bar{z})+ \omega_{1}(z,\bar{z})+  \omega_{2}(z,\bar{z})+... \nonumber
\\
 &=& 4 a \left(g_{0}(z,\bar{z})+a \; g_{1}(z,\bar{z})+a^2  \; g_{2}(z,\bar{z})+...\right),
\end{eqnarray}
where $a=\frac{g^2 N_c}{16 \pi^2 }$,  $\omega_{0}(z,\bar{z})= 4 a \; g_0 (z,\bar{z})$ is the leading-order~(LO) BFKL eigenvalue, $\omega_{1}(z,\bar{z})=  4 a^2 g_1 (z,\bar{z})$ is the next-to-leading-order~(NLO) BFKL eigenvalue and, finally, $\omega_{2}(z,\bar{z})= 4 a^3 g_2 (z,\bar{z})$ is the next-to-next-to-leading-order~(NNLO) BFKL eigenvalue. In the present paper we limit ourselves to the NNLO order, though our analysis can be extended to higher orders as well. 
 
%% testhere
The known LO and NLO eigenvalues in $N=4$ SYM~\cite{NLO,Kotikov:2001sc,Kotikov:2002abmaxtrans}   were  originally derived in terms of polygamma functions and their generalizations and read
 
\begin{eqnarray}\label{g0}
g_0 (z,\bar{z})=2 \left(2 \psi(1) -\psi(z+1)-\psi(\bar{z}+1)  \right)
\end{eqnarray}
and 
\begin{eqnarray}\label{g1}
g_1 (z,\bar{z})= \Phi(z+1) +\Phi(\bar{z}+1)- g_0 (z,\bar{z}) \left(\beta'(z+1)+\beta'(\bar{z}+1)+\frac{\pi^2}{6}\right), \;\; 
\end{eqnarray}
where the functions $\Phi(z)$ and $\beta'(z)$ are given by 
\begin{eqnarray}\label{beta}
\beta'(z)=\sum_{r=0}^{\infty} \frac{(-1)^{r+1}}{(z+r)^2}
\end{eqnarray}
and 
\begin{eqnarray}\label{phi}
\Phi(z)= 3 \zeta(3) +\psi^{''} (z) +2 \Phi_2(z)+2 \beta'(z) \left(\psi(1)-\psi(z)\right)
\end{eqnarray}
in terms of  $\Phi_2(z)$  defined as follows 
\begin{eqnarray}
\Phi_2(z)= \sum_{k=0}^{\infty} \frac{\beta'(k+1)+ (-1)^k \psi'(k+1)}{k+z}-\sum_{k=0}^{\infty}\frac{(-1)^k (\psi(k+1)-\psi(1))}{(k+z)^2}. \;\;\;
\end{eqnarray}

All functions building eq.~(\ref{g0}) and eq.~(\ref{g1}) are meramorphic functions that have the same pole structure, namely, all of them have poles at negative integer values of the argument, where the maximal pole order is limited by the order of the perturbative expansion, simple pole in eq.~(\ref{g0}) for one loop and third order pole in eq.~(\ref{g1}) for two loops.  In general, for any given order $L$ of the perturbative expansion the maximal order of the poles of the functions building the corresponding BFKL eigenvalue is given by $2L-1$. For example, at the three loop perturbative order~(NNLO eigenvalue) all functions are limited to the fifth order pole at most. This fact is related to the 
\emph{maximal transcendentality principle} formulated by Kotikov and Lipatov~\cite{Kotikov:2001sc,Kotikov:2002abmaxtrans} for anomalous dimension for twist-2 operators and restricts the complexity of the underlying functions. Despite the fact that this statement is still lacking a   rigorous proof,  it is natural that at any given loop order of the perturbative expansion the number of the nested integrations is limited by the loop order itself.  As a result this limits the complexity of the resulting functions and their pole structure.  

The leading order~(LO) and the next-to-leading~(NLO) order BFKL eigenvalues in $N=4$ SYM  given in eq.~(\ref{g0}) and eq.~(\ref{g1}) can written in terms of harmonic sums as follows 
 \begin{eqnarray}\label{g0s1}
g_0(z,\bar{z})= -2 \left( S_1(z)+S_1(\bar{z}) \right)
\end{eqnarray}
and 
 \begin{eqnarray}\label{g1s1}
g_1(z,\bar{z})= \Phi(z+1) +\Phi(\bar{z}+1) - 2
 \left[S_1(z)+S_1(\bar{z})\right]
  \left[S_{-2}(z)+S_{-2}(\bar{z})\right],
\end{eqnarray}
where the functions $\Phi(z)$  defined in eq.~(\ref{phi}) in terms of the harmonic sums reads 
\begin{eqnarray}\label{phis1}
\Phi(z+1)&=&
 4 S_{1,-2}(z)-2 S_{-3}(z)  +2 S_3(z)+\frac{\pi^2}{3} S_1 (z)
 \\
 &=& -4 S_{-2,1}(z)+2 S_{-3}(z)+\frac{\pi^2}{3}  S_1(z)+4 S_{-2}(z) S_{1}(z)
 +2 S_3(z) \nonumber
\end{eqnarray}
In eq.~(\ref{g1s1}) and eq.~(\ref{phis1}) the harmonic sums are analytically continued from even integer values of the argument to the complex plane. The analytic continuation from odd integers results into a slightly different expressions, which are beyond the scope of the present paper.  In the next section we use the reflection identities for harmonic sums  to show how  the cross products $S_{\{\alpha\}}(z)  S_{\{\beta\}}(\bar{z})$ can be pole  decomposed into two separate pieces $F(z)$  and $F(\bar{z})$ for $n=0$ and $n=1$, reproducing the corresponding results of  Gromov, Levkovich-Maslyuk and Sizov~\cite{GROMOV},Caron-Huot and Herranen~\cite{HUOT} and Caron-Huot~\cite{Caron-Huot:2015bja}
.

\section{Pole Decomposition of BFKL eigenvalue}

In this section we  demonstrate how the pole decomposition of the BFKL eiegenvalue can be  done using the reflection identities of harmonic sums derived by the authors~\cite{refl2,refl4,refl5}.

The NLO BFKL eigenvalue in $N=4$ SYM in the  color singlet channel is given by eq.~(\ref{g1}). It can be written in terms of the analytically continued harmonic sums as follows~(see eq.~(\ref{g1s1}))
 \begin{eqnarray}\label{g1s2}
g_1(z,\bar{z})= \Phi(z+1) +\Phi(\bar{z}+1) - 2
 \left[S_1(z)+S_1(\bar{z})\right]
  \left[S_{-2}(z)+S_{-2}(\bar{z})\right]
\end{eqnarray}
where the functions $\Phi(z)$ defined in eq.~(\ref{phi}) in terms of the harmonic sums reads 
\begin{eqnarray}\label{phis2}
\Phi(z+1)&=&
 4 S_{1,-2}(z)-2 S_{-3}(z)  +2 S_3(z)+\frac{\pi^2}{3} S_1 (z)
 \\
 &=& -4 S_{-2,1}(z)+2 S_{-3}(z)+\frac{\pi^2}{3}  S_1(z)+4 S_{-2}(z) S_{1}(z)
 +2 S_3(z) \nonumber
\end{eqnarray}
The first line of eq.~(\ref{phis2}) is in  so-called linear basis, whereas the second line is presented in the  non-linear functional basis. Both linear and non-linear basis are equivalent and the transition between the two can be easily done using quasi-shuffle identities of the harmonic sums. The quasi shuffle identities are implemented in \textit{Harmonic Sums} package~\cite{HarmSum}.  The linear basis allows to use the reflection identities in a more  efficient way, this is the major reason for choosing    the linear basis for our analysis.  Only the last term has the mixed pole structure due to the  product of two functions: one of argument $z$ and the another one of argument $\bar{z}$, namely
\begin{eqnarray}
  S_{1}(z) S_{-2}(\bar{z}) + S_{1}(\bar{z}) S_{-2}(z). 
\end{eqnarray}
Let us take a close look at $S_{1}(z) S_{-2}(\bar{z})$ for $n=0$, where $z$ and $\bar{z}$ are not independent anymore and related by $\bar{z}=-1-z$.  The expression 
\begin{eqnarray}
S_{1}(z) S_{-2}(-1-z)
\end{eqnarray}
has poles at both positive and negative integers as well as at zero, while $S_{1}(z)$ has poles only at negative integer values of $z$
 and  $S_{-2}(-1-z)$ has poles at non-negative integer values of $z$. 
 We use the reflection identity~\cite{refl2}
 \begin{eqnarray}\label{s1zsm2m1mz}
&&  S_1(z) S_{-2}(-1-z) = -\frac{1}{2} \pi ^2 \log (2)+\frac{3 \zeta (3)}{4}-\frac{1}{4} \pi ^2
   S_{-1}(-1-z)
   -\frac{1}{4} \pi ^2 S_{-1}(z)
   \nonumber
    \\
 && 
 \hspace{1.9cm}+\frac{1}{12} \pi ^2 S_1(-1-z)-\frac{1}{12} \pi ^2
   S_1(z)+S_{-2,1}(z)+S_{1,-2}(-1-z)
\end{eqnarray}
  to decompose the product into two pieces, one with poles in the left  semi-plane and the another one in the right semi-plane of the complex $z$ plane. Adding the symmetric expression $ S_{1}(\bar{z}) S_{-2}(z)$ we write the whole cross term as follows  
  \begin{eqnarray}
  &&S_{1}(z) S_{-2}(-1-z) + S_{1}(-1-z) S_{-2}(z)= 
  -\pi^2 \ln 2 +\frac{3}{2} \zeta(3)
  \nonumber
   \\
  &&
   -\frac{\pi^2}{2} \left(S_{-1} (z)+S_{-1} (-1-z)\right) 
  +S_{-2,1} (z)+S_{-2,1} (-1-z) 
  \nonumber
   \\
  &&+S_{1,-2} (z)+S_{1,-2} (-1-z) 
\end{eqnarray}
Plugging this into eq.~(\ref{g1s2}) and using the quasi-shuffle identity  for harmonic sums 
\begin{eqnarray}\label{quasis1m2}
S_{1,-2}(z)= S_{-3}(z)+S_{-2}(z) S_{1}(z)-S_{-2,1}(z)
\end{eqnarray}
we obtain the full expression for the NLO BFKL singlet eigenvalue in $N=4$ SYM for $n=0$ 
\begin{eqnarray}
g_1(z,-1-z)&=&-4 S_{-2,1}(z)-4 S_{-2,1}(-1-z)+\pi ^2 S_{-1}(z)
+\frac{\pi^2}{3} ^2 S_{1}(z) \nonumber
\\
&& 
+2 S_{3}(z)
+\pi ^2 S_{-1}(-1-z)+\frac{\pi^2}{3}   S_{1}(-1-z) \nonumber
\\
&&
+2 S_{3}(-1-z)-3 \zeta (3)+2 \pi ^2 \ln 2
\end{eqnarray}
This expression is manifestly symmetrical with respect to $z \leftrightarrow \bar{z}$ and  can be compactly written as a sum of two terms
\begin{eqnarray}\label{deltasn0}
g_1(z,\bar{z})^{n=0}=F^{n=0}_2(z)+F^{n=0}_2(\bar{z}),
\end{eqnarray}
where the function $F^{n=0}_2(z)$   reproduces the function introduced by   Gromov, Levkovich-Maslyuk and Sizov~\cite{GROMOV} (see  eq.~(4) in their paper ) and it is given by 
\begin{eqnarray}\label{F2n0}
F^{n=0}_2(z)= -\frac{3}{2} \zeta(3)+\pi^2 \ln 2+\frac{\pi^2 }{3} S_1(z)+2 S_3(z)+\pi^2 S_{-1}(z) -4 S_{-2,1}(z) \;\; 
\end{eqnarray}

It should be emphasized that we do not  present an alternative way to that of  Gromov, Levkovich-Maslyuk and Sizov~\cite{GROMOV} who calculated this expression by direct pole decomposition.  Instead, we show that the reflection identities based on the techniques used by in Ref.~\cite{GROMOV} can give clear picture of separating the whole expression into two equal pieces for any given value of the conformal spin $n$. It also works in the opposite direction, namely,  the power of reflection identities is in restoring the full functional dependence on $z$ and $\bar{z}$ from the decomposed expression in eq.~(\ref{deltasn0}). In going back from eq.~(\ref{deltasn0}) to the full NLO expression in eq.~(\ref{g1}) we need to consider all possible cross terms at weight three of the type $S_{\{\alpha\}}(z)S_{\{\beta\}}(\bar{z})$ and then decompose them using the reflection identities calculated by one of the authors in Ref.~\cite{refl2}.

In a similar way we decompose the NLO  BFKL eigenvalue in  eq.~(\ref{g1}) for $n=1$ as follows
\begin{eqnarray}\label{deltasn1}
g_1(z,\bar{z})^{n=1}=F^{n=1}_2(z)+F^{n=1}_2(\bar{z}),
\end{eqnarray}
where $F_2^{n=1}(z)$  is calculated  using the reflection identities for harmonic sums~(see Appendix A) and reads
\begin{eqnarray}\label{AF2n1new}
F_2^{n=1}(z)=-\frac{2 S_1(z)}{z^2}+\frac{2 S_{-2}(z)}{z}+\frac{1}{3} \pi ^2 S_1(z)+2 S_3(z)+\zeta (3)
\end{eqnarray}
The expression in eq.~(\ref{deltasn1}) reproduces result of Caron-Huot and Herranen~\cite{HUOT}
  of the NLO BFKL eigenvalue for  $n=1$~(see eq.~(C4) of their paper). 

The full expression for the BFKL eigenvalue at any order is   a functions of only two variables $z=-\frac{1}{2}+\frac{i \nu }{2}+\frac{n}{2}$ and $\bar{z}-\frac{1}{2}-\frac{i \nu }{2}+\frac{n}{2}$. It is useful to consider two cases of either $z +\bar{z}= -1+n$ or $z-\bar{z}=i \nu $ kept fixed. In the present study we focus on $z +\bar{z}= -1+n$ case for specific values of $n$, namely $n=0$ and $n=1$ for which $\bar{z}=-1-z$ and $\bar{z}=-z$ respectively. In this case it natural to reduce two dimensional problem of $z$ and $\bar{z}$ dependence to a problem where the full expression depends on either only $z$ or $\bar{z}$. This rather simple statement regarding any representation similar to that of  eq.~(\ref{F2n0})  allows us to reformulate the problem of finding the full expression for the BFKL eigenvalue in the following way. Our  claim is  that knowing $F^{n}_L(z)$  at given loop order $L$ for any two specific values of $n$  is enough to restore the full functional dependence of the BFKL eigenvalue in $N=4$ SYM at that  loop order.  The additional information one should have is about the space of functions and the relations between the cross terms of the type 
$S_{\{\alpha\}}(z)S_{\{\beta\}}(\bar{z})$ for those two specific values of $n$. Fortunately,  the harmonic sums build the complete basis of functions of one variable~\footnote{It is worth mentioning that the harmonic sums form over-complete basis due to the quasi shuffle identities in eq.~(\ref{shuffle}) between them. The redundancy is removed by choosing either linear or non-linear basis. Our analysis is based on linear basis, which does not include bilinear, trilinear etc. terms of harmonic sums of the same argument. } and the reflection identities provide all possible relations between the cross terms for any two specific values of $n$. The reflections identities for $n=0$ are all known up to weight five  needed for thee loop eigenvalue, while the complete set of them for $n=1$ is still to be calculated. Our statement is based on the factorization  assumption that any  function of two   variables $z$ and $\bar{z}$  in the  
  eigenvalue can be written in terms of a product of two  harmonic sums  $S_{\{\alpha\}}(z)$ and  $S_{\{\beta\}}(\bar{z})$. The factorization assumption follows from the holomorphic separability property of the BFKL eigenvalue, which in its turn originates from the Bethe-Salpeter approach to the BFKL evolution introduced by Kotikov and Lipatov~\cite{Kotikov:2001sc,Kotikov:2002abmaxtrans}(see also Ref.~\cite{BETHE}).  
 
In the next section we show how one can restore the most complicated part of the NNLO BFKL eigenvalue in the singlet channel in $N=4$ SYM using the factorization assumption and the reflection identities of the harmonic sums.

   \section{Restoring NNLO eigenvalue using shifted reflection identities}

In this section we show how one can restore the most complicated  part of the NNLO BFKL eigenvalue using reflection identities for harmonic sums. 
We  use NNLO expression for $n=0$ by  Gromov, Levkovich-Maslyuk and Sizov~\cite{GROMOV} as a starting point in our analysis.  As it was already mentioned in the previous section it  is not enough to use only the $n=0$ expression for the NNLO eigenvalue and  we need to supplement with some additional  information. The natural choice would be to make use of the  $n=1$ expression for the NNLO eigenvalue by Caron-Huot and Herranen~\cite{HUOT}
 as it was independently derived using a different technique. However, we prefer to use some other  input and leave the results of Caron-Huot and Herranen~\cite{HUOT}
 for cross checking our results. 

The additional information we use comes from the observation that the most complex harmonic sums of the largest depth  appearing in $\Phi(z)$ of eq.~(\ref{phis2}) are canceled against the terms coming from the decomposition of the cross products $S_1(z) S_{-2}(\bar{z})+S_1(\bar{z}) S_{-2}(z)$ for $n=1$ in eq.~(\ref{AF2n1new}). This  conclusion is compatible with the result of Caron-Huot and Herranen~\cite{HUOT}
 for NNLO eigenvalue  for $n=1$~(see eq.~(C.5) of their paper).

As the first step in calculating the most complicated term of the NNLO BFKL eigenvalue we take a close look at the  $n=0$ result of  Gromov, Levkovich-Maslyuk and Sizov~\cite{GROMOV} 
\begin{eqnarray}
&&\frac{F_3}{256}=-\frac{5 s_{-5}}{8} -\frac{s_{-4,1}}{2}+\frac{s_1 s_{-3,1}}{2} +\frac{s_{-3,2}}{2}-\frac{5 s_2 s_{-2,1}}{4}+\frac{s_{-4}s_1}{4}  \nonumber 
\\
&&+\frac{s_{-3}s_2}{8}+\frac{3 s_{3,-2}}{4}-\frac{3 s_{-3,1,1}}{2}-s_1 s_{-2,1,1} +s_{2,-2,1} +3s_{-2,1,1,1}  \nonumber
\\
&& 
 -\frac{3 s_{-2}s_3}{4} -\frac{s_5}{8}+\frac{s_{-2}s_1 s_2}{4}
 + \pi^2 \left(\frac{s_{-2,1}}{8}-\frac{7 s_{-3}}{48}-\frac{s_{-2}s_1}{12}+\frac{s_1 s_2}{48}\right)  \nonumber
 \\
 && -\pi^4 \left(\frac{2 s_{-1}}{45}-\frac{s_1}{96}\right)
 +\zeta_3 \left( -\frac{7 s_{-1,1}}{4} +\frac{7s_{-2}}{8}
 + \frac{7 s_{-1} s_{1}}{4}-\frac{s_2}{16} \right)
 \nonumber
 \\
 && 
 \left(2 \mathtt{Li}_4 \left(\frac{1}{2}\right)-\frac{\pi^2 \ln^2 2}{12} +\frac{\ln^4 2 }{12}\right)\left(s_{-1}-s_1\right)+\frac{\ln^5 2}{60}-\frac{\pi^2 \ln^3 2}{36} \nonumber
 \\
 &&
 -\frac{2 \pi^4 \ln 2 }{45} -\frac{\pi^2 \zeta_3}{24}+\frac{49 \zeta_5}{32}-2  \mathtt{Li}_5 \left(\frac{1}{2}\right). 
\end{eqnarray}
Here we use a compact notation of $s_{\{\alpha\}}=S_{\{\alpha\}}(z)$ and   $\bar{s}_{\{\alpha\}}=S_{\{\alpha\}}(\bar{z})$ 
It is useful to write  the same expression in the linear basis 
\begin{eqnarray}\label{F3lin}   %%% Checked Math
&&\frac{F_3}{256}= s_{-1} Z_{-3,1}-s_1 Z_{-3,1}
+\frac{7 \zeta_3  s_{1,-1}}{4} +\frac{s_{-4,1}}{4}
 -\frac{s_{-3,2}}{8} +\frac{\pi ^2 s_{-2,1}}{24} \nonumber
 \\
 && +\frac{s_{-2,3}}{4} -\frac{\pi ^2 s_{1,-2}}{12} 
 -\frac{ s_{2,-3}}{8}-\frac{s_{3,-2}}{4} 
 +\frac{s_{-3,1,1}}{2} +\frac{ s_{1,-3,1}}{2}
 +\frac{ s_{1,-2,2}}{4} \nonumber
 \\
 &&
 +\frac{s_{1,2,-2}}{4} 
 +\frac{s_{2,1,-2}}{4} -s_{1,-2,1,1}+Z_{-3,1,1}+\ln 2\; Z_{-3,1}-\frac{7 s_{-2} \zeta_3}{8}  \nonumber
 \\
 &&
 -\frac{7  s_{-1}}{4} \zeta_3   \ln  2+\frac{7  s_1}{4} \zeta_3   \ln 2  -\frac{\pi ^2  s_{-3}}{16} -\frac{17 \pi ^4 s_{-1}}{720} -\frac{\pi ^4 s_1}{96} +\frac{  \pi ^2 s_1   s_2}{48} \nonumber
 \\
 &&
 -\frac{s_5}{8}-\frac{\pi ^2 \zeta_3 }{8}
 +\frac{\zeta_5 }{2}-\frac{  7 \zeta_3 }{8}  \ln^2 2-\frac{17 \pi^4 \ln 2 }{720}   -\frac{ s_2 \zeta_3  }{16},
 \end{eqnarray}
The constants $Z_{-3,1} $ and $Z_{-3,1,1}$  in eq.~(\ref{F3lin}) are given by 
\begin{eqnarray} %%% Checked Math
Z_{-3,1}=2 \; \mathtt{Li}_4\left(\frac{1}{2}\right)
+\frac{7}{4} \zeta_3 \ln 2-\frac{\pi ^4}{48}+\frac{\ln^4 2}{12}-\frac{\pi^2 \ln^2 2}{12}\simeq 0.0877857 
\end{eqnarray} 

\begin{eqnarray} %%% Checked Math
  Z_{-3,1,1}&=&-2 \mathtt{Li}_5 \left(\frac{1}{2}\right)-2 \mathtt{Li}_4\left(\frac{1}{2}\right) \ln 2+\frac{\pi ^2 \zeta_3}{12}
+\frac{33 \zeta_5}{32} \nonumber
\\
&& -\frac{7}{8} \zeta_3 \ln^2 2-\frac{ \ln^5 2}{15}
+\frac{\pi ^2 \ln^3 2}{18}  \simeq -0.00960157
\end{eqnarray}

We can sort the terms in eq.~(\ref{F3lin}) by complexity of the harmonic sums, i.e. by largest depth for a given weight. For example, 
the sums $s_{-1,1,1,1,1}$ would be one of the most complicated functions because it is of the highest weight $w=5$ and the largest depth $d=5$, but it is absent in eq.~(\ref{F3lin}) as well as any  other sum of depth $d=5$. It follows directly from the definition of the harmonic sums in eq.~(\ref{defS}) that the maximal depth $d$ for a given sum is limited by its  weight $w$ according to $d \le w$. At weight $w=5$  and depth $d=4$ there is only one sum $s_{1,-2,1,1}$ in eq.~(\ref{F3lin}), at weight $w=5$ and depth $d=3$ there are $s_{-3,1,1},s_{1,-3,1},s_{1,-2,2},s_{1,2,-2},s_{2,1,-2}$ etc. All the harmonic sums build the NNLO BFKL eigenvalue for $n=0$ in the linear basis given by in eq.~(\ref{F3lin}) are listed in Tables~\ref{tablew5}-\ref{tablew2}.

\begin{table}[h]
\caption{Harmonic sums of weight $w=5$ appearing in the NNLO expression for $n=0$ of  eq.~(\ref{F3lin}).  }\label{tablew5}
\centering
\begin{tabular}{ |c|c|} 
 \hline
 & weight $5$  \\
   \hline
   depth  5 &  none   \\
   \hline
   depth  4 &  $s_{1,-2,1,1}$   \\
   \hline
   depth  3 &    $s_{-3,1,1},s_{1,-3,1},s_{1,-2,2},s_{1,2,-2},s_{2,1,-2}$  \\
   \hline
   depth  2 &  $s_{-4,1},s_{-3,2},s_{-2,3},s_{2,-3},s_{3,-2}$
    \\
   \hline
   depth  1 &  $s_5$  \\
   \hline
\end{tabular}
\end{table}

\begin{table}[h]
\caption{Harmonic sums of weight $w=4$ appearing in the NNLO expression for $n=0$ of  eq.~(\ref{F3lin}).  }\label{tablew4}
\centering 
\begin{tabular}{ |c|c|} 
 \hline
 & weight $4$  \\
       \hline
   depth  4 &  none  \\
   \hline 
   depth  3 &  none    \\
   \hline
   depth  2 & none  \\
   \hline
   depth  1 & none  \\
   \hline
\end{tabular}
\end{table}

 \begin{table}[h]
\caption{Harmonic sums of weight $w=3$ appearing in the NNLO expression for $n=0$ of  eq.~(\ref{F3lin}).  }\label{tablew3}
\centering
\begin{tabular}{ |c|c|} 
 \hline
 & weight $3$  \\
         \hline 
   depth  3 &  none    \\
   \hline
   depth  2 & $s_{-2,1},s_{1,-2}$ \\
   \hline
   depth  1 & $s_{-3}$ \\
   \hline
\end{tabular}
\end{table}

\begin{table}[h]
\caption{Harmonic sums of weight $w=2$ appearing in the NNLO expression for $n=0$ of  eq.~(\ref{F3lin}).  }\label{tablew2}
\centering
 
\begin{tabular}{ |c|c|} 
 \hline
 & weight $2$  \\
         \hline 
   depth  2 & $s_{1,-1}$ \\
   \hline
   depth  1 & $s_{2},s_{-2}$ \\
   \hline
\end{tabular}
 \end{table}
  
 \begin{table} [h]
\caption{Harmonic sums of weight $w=1$ appearing in the NNLO expression of  eq.~(\ref{F3lin}).  }\label{tablew1}
\centering 
\begin{tabular}{ |c|c|} 
 \hline
 & weight $1$  \\
         \hline 
      depth  1 & $s_{1},s_{-1}$ \\
   \hline
\end{tabular}
\end{table}

  Next, we build an alphabet of possible indices~(letters) of the harmonic sums. We treat separately   the harmonic sums multiplied by the transcendental constant $\pi^2$, $\zeta_3$, $\mathtt{Li}_4\left(\frac{1}{2}\right)$ etc. and those that are not, which we call them the \textit{pure} functions. The pure function appearing  in $F_3$ in eq.~(\ref{F3lin}) are all of weight $w=5$ and they are listed in Table~\ref{tablew5}.  
  We are interested in their alphabet because our goal is to find the most complex part of the NNLO BFKL eigenvalue. From the harmonic sums listed in Table~\ref{tablew5} we draw two major conclusions:  $a)$ index $-1$ is absent for pure functions; $b)$ all the harmonic sums have exactly one negative number in their index, which is most probably related to the cylindrical topology of the singlet BFKL equation as it was discussed by one of the authors in Ref.~\cite{alter}(see also Ref.~\cite{Bondarenko:2015tba}).   Based on those two observations the possible choice for most complicated cross terms is rather limited and is given by     
\begin{eqnarray}\label{crosscompact}
&& s_{1} \bar{s}_{1,1,-2} , \;\; s_{1}  \bar{s}_{1,-2,1},
 \;\; s_{1} \bar{s}_{-2,1,1},
 \\
 &&
  s_{1,1} \bar{s}_{1,-2} , \;\; s_{1,1}  \bar{s}_{-2,1}, 
  \;\; s_{1,1,1}  \bar{s}_{-2}, \nonumber
\end{eqnarray}
where we use a compact notation of $s_{\{\alpha\}}=S_{\{\alpha\}}(z)$ and   $\bar{s}_{\{\alpha\}}=S_{\{\alpha\}}(\bar{z})$.

Symmetrizing  the expressions in eq.~(\ref{crosscompact}) and adding all possible harmonic sums of weight $w=5$ we can build an ansazt for the most complicated  part of the full NNLO BFKL eigenvalue as follows
\begin{eqnarray}\label{complexansatz}
&&A_3(z,\bar{z})= a_1 ( s_{1} \bar{s}_{1,1,-2}+ \bar{s}_{1} s_{1,1,-2})+
a_2 ( s_{1}  \bar{s}_{1,-2,1} + \bar{s}_{1}  s_{1,-2,1}) \nonumber 
\\
&&
+a_3 ( s_{1} \bar{s}_{-2,1,1}+ \bar{s}_{1} s_{-2,1,1})
+a_4 (s_{1,1} \bar{s}_{1,-2}  +\bar{s}_{1,1} s_{1,-2} )
\nonumber 
\\
&&+a_5 ( s_{1,1}  \bar{s}_{-2,1}+ \bar{s}_{1,1}  s_{-2,1})
+a_6 ( s_{1,1,1}  \bar{s}_{-2}+\bar{s}_{1,1,1}  s_{-2})\nonumber 
\\
&& + \phi+\bar{\phi}, 
\end{eqnarray} 
where $a_i$ are the rational coefficients to be fixed and $\phi$  and $\bar{\phi}$ is linear combination~(with rational coefficients) of the most complicated possible harmonic sums of weight $w=5$ of argument $z$ and $\bar{z}$ respectively. Based on the outlined arguments the list of the most complicated harmonic sums reads
\begin{eqnarray}\label{mostcomplicatedS}
&& s_{1,1,1,-2} , \;\; s_{1,1,-2,1}, 
 \;\; s_{1,-2,1,1}, \;\;
  s_{-2,1,1,1} 
\end{eqnarray}
This makes the total number of free coefficients to be fixed in $A_{3}(z, \bar{z})$ in eq.~(\ref{complexansatz}) to be equal to ten. 
The ansatz  $A_{3}(z, \bar{z})$ in eq.~(\ref{complexansatz}) have different pole decomposition for $n=0$ ~($\bar{z}=-1-z$) and $n=1$~($\bar{z}=-z$). We use only the known result for $n=0$ and an assumption that there should be no pure harmonic sums for the $n=1$ pole decomposition of the NNLO eigenvalue. As it was already mentioned this assumption is based on the property of  the next-to-leading expansion for $n=1$ and is compatible with the  NNLO result of Caron-Huot and Herranen~\cite{HUOT}
 for $n=1$. 
Using those two assumptions we fix seven out of ten free coefficients and obtain the NNLO eigenvalue  $g_2$ in eq.~(\ref{gnexp}) for $n=1$ as follows
\begin{eqnarray}\label{g2n1}
g_2(z,-z)&=&\frac{128 s_{1,-2,1}}{z}-\frac{128 s_{-2,1,1}}{z}
+\frac{128 \bar{s}_{1,-2,1}}{(-z)}-\frac{128 \bar{s}_{-2,1,1}}{(-z)}
\nonumber 
\\
&& +  \mathtt{simpler} \;\;\;  \mathtt{terms}
\end{eqnarray}
It should be emphasized that the expression in eq.~(\ref{g2n1}) is completely fixed despite the fact that we are left with three free coefficients in our ansatz  $A_3(z,\bar{z})$  in  eq.~(\ref{complexansatz}). This calculation is based on reflection identities derived by the authors~\cite{refl2, refl4, refl5} and  presented in Appendix D. 
Our result in eq.~(\ref{g2n1}) has a structure similar to that of Ref.~\cite{HUOT}(see eq.~(C.5) of their paper), which reads
\begin{eqnarray}\label{g2n1simon}
&&\frac{64 s_{1,-2,1}}{z}-\frac{128 s_{-2,1,1}}{z}-\frac{128 s_{1,1,-2}}{z}
+\frac{64\bar{s}_{1,-2,1}}{(-z)}-\frac{128 \bar{s}_{-2,1,1}}{(-z)}-\frac{128 \bar{s}_{1,1,-2}}{(-z)}
\nonumber 
\\
&& +  \mathtt{simpler} \;\;\;  \mathtt{terms} 
\end{eqnarray}
This similarity is very encouraging, though the precise connection between the two results requires further analysis that will be presented by us elsewhere. 

The remaining three coefficients in our ansatz  can be  fixed imposing an additional condition of having at most logarithmic divergent of the most complicated terms as $|z| \to \infty$. The condition is more strict than compliance with the asymptotic behavior  of the cusp anomalous dimension, which is automatically satisfied by our result even with the three remaining coefficients left unfixed. This happens  because in fixing the first seven coefficients we used $n=0$ result of  Gromov, Levkovich-Maslyuk and Sizov~\cite{GROMOV}, which is constrained by the  asymptotic expansion   of the cusp anomalous dimension.  The details of fixing the remaining three coefficients of our ansatz are given in Appendix D and the resulting expression built of the most complicated terms    for arbitrary $z$ and $\bar{z}$~(i.e. arbitrary $\nu$ and $n$) reads
\begin{eqnarray} \label{ansatzS}
g_2(z,\bar{z})& =& -128 \left( s_{1} \bar{s}_{1,-2,1} + s_{1}  \bar{s}_{1,-2,1} \right)
+256 \left( s_{1,1,-2,1}+\bar{s}_{1,1,-2,1} \right) \nonumber 
\\
&& +  \mathtt{simpler} \;\;\;  \mathtt{terms}
\end{eqnarray} 
Here we use a compact notation $s_{\{\alpha\}}=S_{\{\alpha\}}(z)$ and $\bar{s}_{\{\alpha\}}=S_{\{\alpha\}}(\bar{z})$.

\section{Conclusions and outlook }\label{sec:concl}
   In this paper we analyzed the analytic properties of the BFKL eigenvalue in the singlet channel in $N=4$ SYM as a functions of two real variables, the discrete  parameter called the conformal spin $n$ and the continuous  $\nu$ related to the anomalous dimension of twist-$2$  operators. It is well known that the BFKL eigenvalue can be written in terms of analytic functions of the complex variable $z=-\frac{1}{2}+i\frac{\nu}{2}+\frac{n}{2}$ and its complex conjugate $\bar{z}=-\frac{1}{2}-i\frac{\nu}{2}+\frac{n}{2}$. We analyze the analytic structure of the known  BFKL eigenvalues expressed in terms of the analytically continued harmonic sums and make use of the fact that all the harmonic sums have the same pole structure, namely they have poles at the negative integer values of the argument. While the physical meaning of  those poles is still to be clarified, one can develop a powerful computational techniques that simplifies the calculations of the BFKL eigenvalue expanding it around the poles.  This technique was originally  introduced by  Gromov, Levkovich-Maslyuk and Sizov~\cite{GROMOV} and our analysis is based on their dedicated Mathematica package.  
   
In the first section we explain how using the reflection identities of harmonic sums the next-to-leading~(NLO) BFKL eigenvalue can be decomposed into two equal pieces~(one is a function of $z$ and the another one is the same function of $\bar{z}$) for specific values of $n$. This way we confirm our technique with known decomposition for $n=0$ and $n=1$  NLO expressions by  Gromov, Levkovich-Maslyuk and Sizov~\cite{GROMOV} and Caron-Huot and Herranen~\cite{HUOT}
 respectively. 

In the next section we analyze the known NNLO BFKL eigenvalue for $n=0$ in $N=4$ SYM by  Gromov, Levkovich-Maslyuk and Sizov~\cite{GROMOV} and write the ansatz for the most coomplicated terms for the case of an arbitrary $n$. Our ansatz has ten free coefficients and we fix seven of them using only  the known $n=0$ expression and assumption that pure harmonic sums~( harmonic sums not multiplied by any non-zero power or argument) are not allowed in the NNLO eigenvalue for $n=1$.  This assumption is compatible with the corresponding result of Caron-Huot and Herranen~\cite{HUOT}
, which has a structure similar but not identical to the one we obtain using our approach.  The close similarity of two results is very encouraging, while the difference between the two may signal that either Bethe–Salpeter approach to the BFKL equation we use is not valid or the result of Caron-Huot and Herranen~\cite{HUOT}
 cannot be expressed using functions of one variable or any cross product of them. In any case this requires further analysis, which will be published by us elsewhere. 

The last three free coefficients in our ansatz for the most complicated term of the NNLO BFKL eiegenvalue in $N=4$ SYM we fix imposing an additional condition of having only simple logarithmic divergence as $|z| \to \infty$ removing all higher powers of the logarithmically  divergent pieces, which should cancel among the most complicated terms in our ansatz.  This condition is more strict than the asymptotic behavior of the cusp anomalous dimension, which is automatically satisfied by our ansatz with seven fixed coefficients, because we fixed them based on the $n=0$  NNLO result of Ref.~\cite{GROMOV}, which was in its turn derived using the asymptotic expansion of the cusp anomalous dimension.   The resulting expression for the most complicated terms of the NNLO BFKL eigenvalue in $N=4$ SYM for arbitrary $\nu$ and $n$ is given in eq.~(\ref{ansatzS}) and  presents the main result of our work.

\newpage

%\section*{Appendix}
\newpage
\renewcommand{\theequation}{A-\arabic{equation}}    
  % redefine the command that creates the equation no.    
  \setcounter{equation}{0}  % reset counter   
    \setcounter{section}{0}  % reset counter   
  \section*{APPENDIX}  % use *-form to suppress numbering
\section*{A. Pole decomposition of NLO eigenvalue for $n=1$}
In this section we present the details of the pole decomposition of the NLO BFKL eigenvalue in $N=4$ SYM in the case of the conformal spin equal unity, i.e. $n=1$.  In our analysis we use the shifted reflection identities for $\bar{z}=-z$ obtained from the reflection identities for $\bar{z}=-1-z$ by shifting the argument of the harmonic sums by unity. For example, we shift the argument of $S_{-2}(z)$  
\begin{eqnarray}
S_{-2}(-1-z)=-S_{-2}(-z)+\frac{1}{z^2}-\frac{\pi^2}{6}
\end{eqnarray}
then multiply it by $S_1(z)$
\begin{eqnarray}
S_1(z) S_{-2}(-z)= -S_1(z)S_{-2}(-1-z) +\frac{S_{1}(z)}{z^2}-\frac{\pi^2}{6} S_{1}(z)
\end{eqnarray}
 and finally  plug it into the reflection identity of eq.~(\ref{s1zsm2m1mz}). The resulting shifted reflection identity reads  
\begin{eqnarray}
S_{-2}(z) S_{1}(-z)& =& -S_{-2,1}(z)-S_{-2,1}(-z)-\frac{S_{-2}(z)}{z}+S_{-3}(z)-\frac{1}{4} \pi ^2 S_{-1}(z)  \nonumber
\\
&&
+\frac{1}{12} \pi ^2 S_{1}(z)+S_{-2}(z) S_{1}(z)+\frac{ S_{1}(-z) }{z^2}+\frac{1}{4}
 \pi ^2 S_{-1}(-z)  \nonumber
 \\
 &&
 -\frac{1}{12} \pi ^2 S_{1}(-z)+\frac{\pi ^2}{6 z}-\frac{\zeta (3)}{2}
\end{eqnarray}
After the symmetrization with respect to $\bar{z} \leftrightarrow z$ we obtain the only cross term of the NLO BFKL in eq.~(\ref{g1s1}) eigenvalue decomposed as follows
\begin{eqnarray}\label{s1sm2mzssm2s1mz}
S_{1}(z) S_{-2}(-z)+ S_{1}(-z) S_{-2}(z) &=& -2 S_{-2,1}(z)-2 
S_{-2,1}(-z)+\frac{S_{1}(z)}{z^2}  \nonumber
\\
&& \hspace{-3cm}
-\frac{S_{-2}(z)}{z}+S_{-3}(z)+S_{-2}(z) S_{1}(z)+\frac{S_{1}(-z)}{z^2}+\frac{S_{-2}(-z)}{z}  \nonumber
\\
&& \hspace{-3cm}
+S_{-3}(-z)+S_{-2}(-z) S_{1}(-z)-\zeta (3)
\end{eqnarray}
Next we shift the argument of the harmonic sums appearing  in the function $\Phi$ in eq.(\ref{phis2}) 
\begin{eqnarray}
S_{-2,1}(-1-z)=-S_{-2,1}(-z)+\frac{S_1(-z)}{z^2}-\frac{5}{4}\zeta(3)
\end{eqnarray}
and  
\begin{eqnarray}
S_{-3}(-1-z)=-S_{-3}(-z)-\frac{1}{z^3}-\frac{3}{2}\zeta(3)
\end{eqnarray}
  as well as  
\begin{eqnarray}
S_{1}(-1-z)=S_{1}(-z)+\frac{1}{z}. 
\end{eqnarray}

\begin{eqnarray}
S_{1,-2}(-1-z)&= & S_{-2,1}(-z)-S_{-3}(-z)-S_{-2}(-z)S_{1}(-z)
\nonumber 
\\
&& -\frac{S_{-2}(-z)}{z}+\frac{3}{2} \zeta(3)-\frac{\pi^2}{6}\frac{1}{z} +\frac{5}{4}\zeta(3)
\end{eqnarray}

\begin{eqnarray}
S_{1,-2}(-1-z)&=&
S_{-2,1}(-z)-\frac{S_{-2}(-z)}{z}-S_{-3}(-z)
\\
&&+\frac{\pi^2}{6} S_1(-z)-S_{-2}(-z) S_1(-z)+\frac{\pi ^2}{6 }\frac{1}{z}+\frac{11 }{4} \zeta (3)  \nonumber
\end{eqnarray}

Finally, using the expressions in eq.~(\ref{s1sm2mzssm2s1mz}) and  eq.~(\ref{phis2})  we  write the BFKL eigenvalue in eq.~(\ref{deltasn1}) as follows 
 \begin{eqnarray}
g_1(z,-z)&=&-\frac{2 S_1(z)}{z^2}+\frac{2 S_{-2}(z)}{z}+\frac{1}{3} \pi ^2 S_1(z)+2 S_3(z)
\\
&&-\frac{2 S_1(-z)}{z^2}
-\frac{2  S_{-2}(-z)}{z}
+\frac{1}{3} \pi ^2  S_1(-z)+2  S_3(-z)+2 \zeta (3)  \nonumber
\\
&=& -\frac{2 S_1(z)}{z^2}+\frac{2 S_{-2}(z)}{z}+\frac{1}{3} \pi ^2 S_1(z)+2 S_3(z) + \zeta (3)\nonumber
\\
&&-\frac{2 S_1(-z)}{(-z)^2}
+\frac{2  S_{-2}(-z)}{(-z)}
+\frac{1}{3} \pi ^2  S_1(-z)+2  S_3(-z)+ \zeta (3) \nonumber,
\end{eqnarray} 
which is symmetrical with respect to $z \leftrightarrow \bar{z}$ for $n=1$. This  allows to write the NLO eigenvalue in a compact way  
\begin{eqnarray}\label{Adeltasn1}
\delta(z,\bar{z})^{n=1}= F_2^{n=1}(z)+F_2^{n=1}(\bar{z})
\end{eqnarray}
 using the  function of one variable 
\begin{eqnarray}\label{AF2n1}
F_2^{n=1}(z)=-\frac{2 S_1(z)}{z^2}+\frac{2 S_{-2}(z)}{z}+\frac{1}{3} \pi ^2 S_1(z)+2 S_3(z)+\zeta (3)
\end{eqnarray}
Note, the function $F_2^{n=1}(z)$  in eq.~(\ref{AF2n1}) is the function obtained by Caron-Huot and Herranen~\cite{HUOT}
 except for overall normalization factor which stems from a different definition of the perturbative expansion.

\section*{B. Pole decomposition of  most complicated cross products at weight five  for $n=0$}
 \renewcommand{\theequation}{B-\arabic{equation}}
  % redefine the command that creates the equation no.
  \setcounter{equation}{0}  % reset counter 
  
  In this section  we present the pole decomposition of  cross products 
 $S_{\{\alpha\}} (z) S_{\{\beta\}}(\bar{z})$ for $n=0$ i.e. for $\bar{z}=-1-z$. All the relevant cross products are presented in the compact notation $S_{\{\alpha\}}(z)=s_{\{\alpha\}}$, $S_{\{\alpha\}}(\bar{z})={\bar{s}}_{\{\alpha\}}$, $\ln(2) =\ln_2$, $\zeta(n)=\zeta_n$ and $\mathtt{Li}_n= \mathtt{Li}_n \left(\frac{1}{2}\right)$, where $\zeta(n)$ is the Riemann zeta function and $\mathtt{Li}_n \left(\frac{1}{2}\right)$ is the polylogarithm of argument one half. The transcendental constants appearing here together with $\pi$ build the irreducible set of the transcendental  constants, which was used to construct the functional basis for the pole expansion.  Each   cross product $S_{\{\alpha\}} (z) S_{\{\beta\}}(\bar{z})$  is  decomposed into two functions, one of $z$ and another one of $\bar{z}$. Below each decomposition we write an expression that contains only the most complicated functions building it. In the present study we focus only on the most complicated functions in deriving the most complicated part of the NNLO BFKL eigenvalue for arbitrary $\nu$ and $n$. The reflection identities listed here were calculated by the authors in Ref.~\cite{refl5}. 
 
\begin{dmath}
\bar{s}_1 s_{-2,1,1}+s_1 \bar{s}_{-2,1,1}= \bar{s}_{-4,1}+\frac{1}{3} \pi ^2 \bar{s}_{-2,1}-2
   \bar{s}_{-3,1,1}-2 \bar{s}_{-2,2,1}+3 \bar{s}_{-2,1,1,1}+\bar{s}_{1,-2,1,1}-2 \mathtt{Li}_4
   \bar{s}_{-1}+3 \zeta _3 \bar{s}_{-2}-\frac{7}{4} \zeta _3 \ln _2 \bar{s}_{-1}-\frac{1}{6} \pi ^2
   \bar{s}_{-3}-\frac{43}{720} \pi ^4 \bar{s}_{-1}-\frac{1}{12} \ln _2^4 \bar{s}_{-1}+\frac{1}{12} \pi ^2
   \ln _2^2 \bar{s}_{-1}+s_{-4,1}+\frac{1}{3} \pi ^2 s_{-2,1}-2 s_{-3,1,1}-2 s_{-2,2,1}+3
   s_{-2,1,1,1}+s_{1,-2,1,1}+\frac{\pi ^2 \zeta _3}{48}+\frac{7 \zeta _5}{2}-\frac{7}{4} \zeta _3 \ln
   _2^2-2 \mathtt{Li}_4 s_{-1}+4 \mathtt{Li}_5+3 \zeta _3 s_{-2}-\frac{7}{4} \zeta _3 s_{-1} \ln
   _2-\frac{1}{6} \pi ^2 s_{-3}-\frac{43}{720} \pi ^4 s_{-1}-\frac{1}{12} s_{-1} \ln _2^4+\frac{1}{12}
   \pi ^2 s_{-1} \ln _2^2-\frac{\ln _2^5}{30}+\frac{1}{18} \pi ^2 \ln _2^3-\frac{43 \pi ^4 \ln _2}{360}
\end{dmath}
and its the most complicated part is given by 
\begin{dmath}
\bar{s}_1 s_{-2,1,1}+s_1 \bar{s}_{-2,1,1}\to 3 \bar{s}_{-2,1,1,1}+\bar{s}_{1,-2,1,1}+3
   s_{-2,1,1,1}+s_{1,-2,1,1}
\end{dmath}

%%%%%%%%%%%%%%%%%%%%%%%%%%%%%%%%%%%%%%%%%%%%%%%%%%%%%%%%%%%%%%

\begin{dmath}
\bar{s}_1 s_{1,-2,1}+s_1 \bar{s}_{1,-2,1}= -\frac{21}{4} \zeta _3
   \bar{s}_{1,-1}+\bar{s}_{-4,1}+\frac{1}{3} \pi ^2 \bar{s}_{1,-2}-\bar{s}_{-3,1,1}-2
   \bar{s}_{1,-3,1}-\bar{s}_{2,-2,1}+2 \bar{s}_{1,-2,1,1}+2 \bar{s}_{1,1,-2,1}-6 \mathtt{Li}_4
   \bar{s}_{-1}+6 \mathtt{Li}_4 \bar{s}_1+\frac{21}{8} \zeta _3 \bar{s}_{-2}-\frac{5}{8} \zeta _3
   \bar{s}_2-\frac{1}{6} \pi ^2 \bar{s}_{-3}+\frac{1}{120} \pi ^4 \bar{s}_{-1}-\frac{11}{288} \pi ^4
   \bar{s}_1-\frac{1}{4} \ln _2^4 \bar{s}_{-1}+\frac{1}{4} \ln _2^4 \bar{s}_1+\frac{1}{4} \pi ^2 \ln _2^2
   \bar{s}_{-1}-\frac{1}{4} \pi ^2 \ln _2^2 \bar{s}_1-\frac{21}{4} \zeta _3 s_{1,-1}+s_{-4,1}+\frac{1}{3}
   \pi ^2 s_{1,-2}-s_{-3,1,1}-2 s_{1,-3,1}-s_{2,-2,1}+2 s_{1,-2,1,1}+2 s_{1,1,-2,1}-\frac{7 \pi ^2 \zeta
   _3}{12}-\frac{107 \zeta _5}{16}-6 \mathtt{Li}_4 s_{-1}+6 \mathtt{Li}_4 s_1+12 \mathtt{Li}_5+\frac{21}{8}
   \zeta _3 s_{-2}-\frac{5 \zeta _3 s_2}{8}-\frac{1}{6} \pi ^2 s_{-3}+\frac{1}{120} \pi ^4
   s_{-1}-\frac{11 \pi ^4 s_1}{288}-\frac{1}{4} s_{-1} \ln _2^4+\frac{1}{4} s_1 \ln _2^4+\frac{1}{4} \pi
   ^2 s_{-1} \ln _2^2-\frac{1}{4} \pi ^2 s_1 \ln _2^2-\frac{\ln _2^5}{10}+\frac{1}{6} \pi ^2 \ln
   _2^3+\frac{\pi ^4 \ln _2}{60}
\end{dmath}
and its the most complicated part is given by 
\begin{dmath}
\bar{s}_1 s_{1,-2,1}+s_1 \bar{s}_{1,-2,1}\to 2 \bar{s}_{1,-2,1,1}+2 \bar{s}_{1,1,-2,1}+2 s_{1,-2,1,1}+2
   s_{1,1,-2,1}
\end{dmath}

%%%%%%%%%%%%%%%%%%%%%%%%%%%%%%%%%%%%%%%%%%%%%%%%%%%%%%%%%%%%

\begin{dmath}
\bar{s}_1 s_{1,1,-2}+s_1 \bar{s}_{1,1,-2} = -\frac{7}{2} \zeta _3 \bar{s}_{1,-1}+\frac{3}{4} \zeta _3
   \bar{s}_{1,1}+\bar{s}_{-4,1}+\frac{1}{4} \pi ^2 \bar{s}_{1,-2}-\frac{1}{12} \pi ^2
   \bar{s}_{1,2}+\frac{1}{4} \pi ^2 \bar{s}_{2,-1}+\frac{1}{12} \pi ^2
   \bar{s}_{2,1}-\bar{s}_{1,-3,1}-\frac{1}{2} \pi ^2
   \bar{s}_{1,1,-1}-\bar{s}_{1,2,-2}-\bar{s}_{2,-2,1}-\bar{s}_{2,1,-2}+\bar{s}_{1,1,-2,1}+3
   \bar{s}_{1,1,1,-2}+\frac{1}{2} \pi ^2 \ln _2 \bar{s}_{1,-1}-\frac{1}{2} \pi ^2 \ln _2 \bar{s}_{1,1}-2
   \mathtt{Li}_4 \bar{s}_{-1}+4 \mathtt{Li}_4 \bar{s}_1+\frac{7}{4} \zeta _3 \bar{s}_{-2}-\frac{3}{4} \zeta
   _3 \bar{s}_2+\frac{7}{4} \zeta _3 \ln _2 \bar{s}_{-1}-\frac{1}{6} \pi ^2 \bar{s}_{-3}+\frac{1}{360}
   \pi ^4 \bar{s}_{-1}-\frac{7}{120} \pi ^4 \bar{s}_1-\frac{1}{12} \ln _2^4 \bar{s}_{-1}+\frac{1}{6} \ln
   _2^4 \bar{s}_1-\frac{1}{6} \pi ^2 \ln _2^2 \bar{s}_{-1}+\frac{1}{12} \pi ^2 \ln _2^2
   \bar{s}_1-\frac{1}{4} \pi ^2 \ln _2 \bar{s}_{-2}+\frac{1}{4} \pi ^2 \ln _2 \bar{s}_2-\frac{7}{2} \zeta
   _3 s_{1,-1}+\frac{3}{4} \zeta _3 s_{1,1}+s_{-4,1}+\frac{1}{4} \pi ^2 s_{1,-2}-\frac{1}{12} \pi ^2
   s_{1,2}+\frac{1}{4} \pi ^2 s_{2,-1}+\frac{1}{12} \pi ^2 s_{2,1}-s_{1,-3,1}-\frac{1}{2} \pi ^2
   s_{1,1,-1}-s_{1,2,-2}-s_{2,-2,1}-s_{2,1,-2}+s_{1,1,-2,1}+3 s_{1,1,1,-2}+\frac{1}{2} \pi ^2 \ln _2
   s_{1,-1}-\frac{1}{2} \pi ^2 \ln _2 s_{1,1}-\frac{3 \pi ^2 \zeta _3}{4}-\frac{19 \zeta
   _5}{16}+\frac{7}{4} \zeta _3 \ln _2^2-2 \mathtt{Li}_4 s_{-1}+4 \mathtt{Li}_4 s_1+4 \mathtt{Li}_5+\frac{7}{4}
   \zeta _3 s_{-2}-\frac{3 \zeta _3 s_2}{4}+\frac{7}{4} \zeta _3 s_{-1} \ln _2-\frac{1}{6} \pi ^2
   s_{-3}+\frac{1}{360} \pi ^4 s_{-1}-\frac{7 \pi ^4 s_1}{120}-\frac{1}{12} s_{-1} \ln _2^4+\frac{1}{6}
   s_1 \ln _2^4-\frac{1}{6} \pi ^2 s_{-1} \ln _2^2+\frac{1}{12} \pi ^2 s_1 \ln _2^2-\frac{1}{4} \pi ^2
   s_{-2} \ln _2+\frac{1}{4} \pi ^2 s_2 \ln _2-\frac{\ln _2^5}{30}-\frac{1}{9} \pi ^2 \ln _2^3+\frac{\pi
   ^4 \ln _2}{180}
\end{dmath}
and its the most complicated part is given by 
\begin{dmath}
\bar{s}_1 s_{1,1,-2}+s_1 \bar{s}_{1,1,-2}\to \bar{s}_{1,1,-2,1}+3 \bar{s}_{1,1,1,-2}+s_{1,1,-2,1}+3
   s_{1,1,1,-2}
\end{dmath}

%%%%%%%%%%%%%%%%%%%%%%%%%%%%%%%%%%%%%%%%%%%%%%%%%%%%%%

\begin{dmath}
s_{1,1} \bar{s}_{-2,1}+s_{-2,1} \bar{s}_{1,1} = -\frac{21}{4} \zeta _3 \bar{s}_{1,-1}+2
   \bar{s}_{-4,1}+\frac{1}{3} \pi ^2 \bar{s}_{-2,1}+\frac{1}{3} \pi ^2 \bar{s}_{1,-2}-3
   \bar{s}_{-3,1,1}-2 \bar{s}_{-2,2,1}-2 \bar{s}_{1,-3,1}-\bar{s}_{2,-2,1}+3 \bar{s}_{-2,1,1,1}+2
   \bar{s}_{1,-2,1,1}+\bar{s}_{1,1,-2,1}-6 \mathtt{Li}_4 \bar{s}_{-1}+6 \mathtt{Li}_4 \bar{s}_1+\frac{45}{8}
   \zeta _3 \bar{s}_{-2}-\frac{5}{8} \zeta _3 \bar{s}_2-\frac{1}{3} \pi ^2 \bar{s}_{-3}-\frac{13}{240}
   \pi ^4 \bar{s}_{-1}-\frac{11}{288} \pi ^4 \bar{s}_1-\frac{1}{4} \ln _2^4 \bar{s}_{-1}+\frac{1}{4} \ln
   _2^4 \bar{s}_1+\frac{1}{4} \pi ^2 \ln _2^2 \bar{s}_{-1}-\frac{1}{4} \pi ^2 \ln _2^2
   \bar{s}_1-\frac{21}{4} \zeta _3 s_{1,-1}+2 s_{-4,1}+\frac{1}{3} \pi ^2 s_{-2,1}+\frac{1}{3} \pi ^2
   s_{1,-2}-3 s_{-3,1,1}-2 s_{-2,2,1}-2 s_{1,-3,1}-s_{2,-2,1}+3 s_{-2,1,1,1}+2
   s_{1,-2,1,1}+s_{1,1,-2,1}-\frac{11 \zeta _5}{8}-6 \mathtt{Li}_4 s_{-1}+6 \mathtt{Li}_4 s_1+12
   \mathtt{Li}_5+\frac{45}{8} \zeta _3 s_{-2}-\frac{5 \zeta _3 s_2}{8}-\frac{1}{3} \pi ^2
   s_{-3}-\frac{13}{240} \pi ^4 s_{-1}-\frac{11 \pi ^4 s_1}{288}-\frac{1}{4} s_{-1} \ln _2^4+\frac{1}{4}
   s_1 \ln _2^4+\frac{1}{4} \pi ^2 s_{-1} \ln _2^2-\frac{1}{4} \pi ^2 s_1 \ln _2^2-\frac{\ln
   _2^5}{10}+\frac{1}{6} \pi ^2 \ln _2^3-\frac{13 \pi ^4 \ln _2}{120}
\end{dmath}
and its the most complicated part is given by 
 \begin{dmath}
 s_{1,1} \bar{s}_{-2,1}+s_{-2,1} \bar{s}_{1,1}\to 3 \bar{s}_{-2,1,1,1}+2
   \bar{s}_{1,-2,1,1}+\bar{s}_{1,1,-2,1}+3 s_{-2,1,1,1}+2 s_{1,-2,1,1}+s_{1,1,-2,1}
 \end{dmath}

 %%%%%%%%%%%%%%%%%%%%%%%%%%%%%%%%%%%%%%%%%%%%%%%%%%%%%%%%%%%%%%%%%%%%

 \begin{dmath}
 s_{1,1} \bar{s}_{1,-2}+s_{1,-2} \bar{s}_{1,1}= -\frac{35}{4} \zeta _3 \bar{s}_{1,-1}+\frac{3}{2} \zeta
   _3 \bar{s}_{1,1}+\bar{s}_{-4,1}+\frac{1}{2} \pi ^2 \bar{s}_{1,-2}-\frac{1}{6} \pi ^2
   \bar{s}_{1,2}+\frac{1}{2} \pi ^2 \bar{s}_{2,-1}+\bar{s}_{3,-2}-\bar{s}_{-3,1,1}-2 \bar{s}_{1,-3,1}-\pi
   ^2 \bar{s}_{1,1,-1}-2 \bar{s}_{1,2,-2}-\bar{s}_{2,-2,1}-2 \bar{s}_{2,1,-2}+\bar{s}_{1,-2,1,1}+2
   \bar{s}_{1,1,-2,1}+3 \bar{s}_{1,1,1,-2}+\pi ^2 \ln _2 \bar{s}_{1,-1}-\pi ^2 \ln _2 \bar{s}_{1,1}-10
   \mathtt{Li}_4 \bar{s}_{-1}+10 \mathtt{Li}_4 \bar{s}_1+\frac{35}{8} \zeta _3 \bar{s}_{-2}-\frac{7}{8} \zeta
   _3 \bar{s}_2-\frac{1}{4} \pi ^2 \bar{s}_{-3}+\frac{1}{18} \pi ^4 \bar{s}_{-1}-\frac{149 \pi ^4
   \bar{s}_1}{1440}+\frac{1}{12} \pi ^2 \bar{s}_3-\frac{5}{12} \ln _2^4 \bar{s}_{-1}+\frac{5}{12} \ln
   _2^4 \bar{s}_1-\frac{1}{12} \pi ^2 \ln _2^2 \bar{s}_{-1}+\frac{1}{12} \pi ^2 \ln _2^2
   \bar{s}_1-\frac{1}{2} \pi ^2 \ln _2 \bar{s}_{-2}+\frac{1}{2} \pi ^2 \ln _2 \bar{s}_2-\frac{35}{4}
   \zeta _3 s_{1,-1}+\frac{3}{2} \zeta _3 s_{1,1}+s_{-4,1}+\frac{1}{2} \pi ^2 s_{1,-2}-\frac{1}{6} \pi ^2
   s_{1,2}+\frac{1}{2} \pi ^2 s_{2,-1}+s_{3,-2}-s_{-3,1,1}-2 s_{1,-3,1}-\pi ^2 s_{1,1,-1}-2
   s_{1,2,-2}-s_{2,-2,1}-2 s_{2,1,-2}+s_{1,-2,1,1}+2 s_{1,1,-2,1}+3 s_{1,1,1,-2}+\pi ^2 \ln _2
   s_{1,-1}-\pi ^2 \ln _2 s_{1,1}-\pi ^2 \zeta _3-\frac{185 \zeta _5}{16}-10 \mathtt{Li}_4 s_{-1}+10
   \mathtt{Li}_4 s_1+20 \mathtt{Li}_5+\frac{35}{8} \zeta _3 s_{-2}-\frac{7 \zeta _3 s_2}{8}-\frac{1}{4} \pi
   ^2 s_{-3}+\frac{1}{18} \pi ^4 s_{-1}-\frac{149 \pi ^4 s_1}{1440}+\frac{\pi ^2 s_3}{12}-\frac{5}{12}
   s_{-1} \ln _2^4+\frac{5}{12} s_1 \ln _2^4-\frac{1}{12} \pi ^2 s_{-1} \ln _2^2+\frac{1}{12} \pi ^2 s_1
   \ln _2^2-\frac{1}{2} \pi ^2 s_{-2} \ln _2+\frac{1}{2} \pi ^2 s_2 \ln _2-\frac{\ln
   _2^5}{6}-\frac{1}{18} \pi ^2 \ln _2^3+\frac{\pi ^4 \ln _2}{9}
 \end{dmath}
and its the most complicated part is given by 
 \begin{dmath}
 s_{1,1} \bar{s}_{1,-2}+s_{1,-2} \bar{s}_{1,1}\to \bar{s}_{1,-2,1,1}+2 \bar{s}_{1,1,-2,1}+3
   \bar{s}_{1,1,1,-2}+s_{1,-2,1,1}+2 s_{1,1,-2,1}+3 s_{1,1,1,-2}
 \end{dmath}

 %%%%%%%%%%%%%%%%%%%%%%%%%%%%%%%%%%%%%%%%%%%%%%%%%%%%%
 
 \begin{dmath}
 \bar{s}_{-2} s_{1,1,1}+s_{-2} \bar{s}_{1,1,1} = -\frac{21}{4} \zeta _3 \bar{s}_{1,-1}+\frac{3}{4} \zeta
   _3 \bar{s}_{1,1}+\frac{1}{4} \pi ^2 \bar{s}_{1,-2}-\frac{1}{12} \pi ^2 \bar{s}_{1,2}+\frac{1}{4} \pi
   ^2 \bar{s}_{2,-1}-\frac{1}{12} \pi ^2
   \bar{s}_{2,1}+\bar{s}_{3,-2}-\bar{s}_{-3,1,1}-\bar{s}_{1,-3,1}-\frac{1}{2} \pi ^2
   \bar{s}_{1,1,-1}-\bar{s}_{1,2,-2}-\bar{s}_{2,1,-2}+\bar{s}_{-2,1,1,1}+\bar{s}_{1,-2,1,1}+\bar{s}_{1,1,
   -2,1}+\bar{s}_{1,1,1,-2}+\frac{1}{2} \pi ^2 \ln _2 \bar{s}_{1,-1}-\frac{1}{2} \pi ^2 \ln _2
   \bar{s}_{1,1}-6 \mathtt{Li}_4 \bar{s}_{-1}+6 \mathtt{Li}_4 \bar{s}_1+\frac{21}{8} \zeta _3
   \bar{s}_{-2}-\frac{1}{8} \zeta _3 \bar{s}_2-\frac{1}{12} \pi ^2 \bar{s}_{-3}+\frac{1}{120} \pi ^4
   \bar{s}_{-1}-\frac{13}{288} \pi ^4 \bar{s}_1+\frac{1}{12} \pi ^2 \bar{s}_3-\frac{1}{4} \ln _2^4
   \bar{s}_{-1}+\frac{1}{4} \ln _2^4 \bar{s}_1-\frac{1}{4} \pi ^2 \ln _2 \bar{s}_{-2}+\frac{1}{4} \pi ^2
   \ln _2 \bar{s}_2-\frac{21}{4} \zeta _3 s_{1,-1}+\frac{3}{4} \zeta _3 s_{1,1}+\frac{1}{4} \pi ^2
   s_{1,-2}-\frac{1}{12} \pi ^2 s_{1,2}+\frac{1}{4} \pi ^2 s_{2,-1}-\frac{1}{12} \pi ^2
   s_{2,1}+s_{3,-2}-s_{-3,1,1}-s_{1,-3,1}-\frac{1}{2} \pi ^2
   s_{1,1,-1}-s_{1,2,-2}-s_{2,1,-2}+s_{-2,1,1,1}+s_{1,-2,1,1}+s_{1,1,-2,1}+s_{1,1,1,-2}+\frac{1}{2} \pi
   ^2 \ln _2 s_{1,-1}-\frac{1}{2} \pi ^2 \ln _2 s_{1,1}-\frac{3 \pi ^2 \zeta _3}{16}-\frac{107 \zeta
   _5}{16}-6 \mathtt{Li}_4 s_{-1}+6 \mathtt{Li}_4 s_1+12 \mathtt{Li}_5+\frac{21}{8} \zeta _3 s_{-2}-\frac{\zeta
   _3 s_2}{8}-\frac{1}{12} \pi ^2 s_{-3}+\frac{1}{120} \pi ^4 s_{-1}-\frac{13 \pi ^4 s_1}{288}+\frac{\pi
   ^2 s_3}{12}-\frac{1}{4} s_{-1} \ln _2^4+\frac{1}{4} s_1 \ln _2^4-\frac{1}{4} \pi ^2 s_{-2} \ln
   _2+\frac{1}{4} \pi ^2 s_2 \ln _2-\frac{\ln _2^5}{10}+\frac{\pi ^4 \ln _2}{60}
 \end{dmath}
and its the most complicated part is given by 
 \begin{dmath}
 \bar{s}_{-2} s_{1,1,1}+s_{-2} \bar{s}_{1,1,1}\to
   \bar{s}_{-2,1,1,1}+\bar{s}_{1,-2,1,1}+\bar{s}_{1,1,-2,1}+\bar{s}_{1,1,1,-2}+s_{-2,1,1,1}+s_{1,-2,1,1}+
   s_{1,1,-2,1}+s_{1,1,1,-2}
 \end{dmath}
   In the next section we write the  pole decomposition for the cross terms for $n=1$, i.e. when $\bar{z}=-z$. 
   
\newpage
\section*{C. Pole decomposition of most complicated cross products   for $n=1$}\label{Appn1}
 \renewcommand{\theequation}{C-\arabic{equation}}
  % redefine the command that creates the equation no.
  \setcounter{equation}{0}  % reset counter 

    In this section  we present the pole decomposition of  cross products 
 $S_{\{\alpha\}} (z) S_{\{\beta\}}(\bar{z})$ for $n=1$ i.e. for $\bar{z}=-z$. All the relevant cross products are presented in the compact notation $S_{\{\alpha\}}(z)=s_{\{\alpha\}}$, $S_{\{\alpha\}}(\bar{z})={\bar{s}}_{\{\alpha\}}$, $\ln(2) =\ln_2$, $\zeta(n)=\zeta_n$ and $\mathtt{Li}_n= \mathtt{Li}_n \left(\frac{1}{2}\right)$, where $\zeta(n)$ is the Riemann zeta function and $\mathtt{Li}_n \left(\frac{1}{2}\right)$ is the polylogarithm of argument one half. The transcendental constants appearing here together with $\pi$ build the irreducible set of the transcendental  constants, which was used to construct the functional basis for the pole expansion.  Each   cross product $S_{\{\alpha\}} (z) S_{\{\beta\}}(\bar{z})$  is  decomposed into two functions, one of $z$ and another one of $\bar{z}$. Below each decomposition we write an expression that contains only the most complicated functions building it. In the present study we focus only on the most complicated functions in deriving the most complicated part of the NNLO BFKL eigenvalue for arbitrary $\nu$ and $n$. The reflection identities listed here were calculated by the authors in Ref.~\cite{refl5}. 

\begin{dmath}
\bar{s}_1 s_{-2,1,1}+s_1 \bar{s}_{-2,1,1} =
   \frac{\bar{s}_{1,1}}{z^3}+\frac{-\bar{s}_{2,1}+\bar{s}_{1,1,1}+\frac{1}{6} \pi ^2 \bar{s}_1+2 \zeta
   _3}{z^2}+\frac{\bar{s}_{-2,1,1}}{z}-\bar{s}_{-4,1}+\bar{s}_{-2,1,1,1}+\bar{s}_{1,-2,1,1}-\zeta _3
   \bar{s}_{-2}+\frac{\bar{s}_1}{z^4}+\frac{1}{6} \pi ^2
   \bar{s}_{-3}-\frac{s_{1,1}}{z^3}+\frac{-s_{2,1}+s_{1,1,1}+2 \zeta _3+\frac{\pi ^2
   s_1}{6}}{z^2}-\frac{s_{-2,1,1}}{z}-s_{-4,1}+s_{-2,1,1,1}+s_{1,-2,1,1}+\frac{19 \pi ^2 \zeta
   _3}{48}-\frac{57 \zeta _5}{16}-\zeta _3 s_{-2}+\frac{s_1}{z^4}+\frac{1}{6} \pi ^2 s_{-3}
\end{dmath}
and its the most complicated part is given by 
\begin{dmath}
\bar{s}_1 s_{-2,1,1}+s_1 \bar{s}_{-2,1,1}\to
   \frac{\bar{s}_{-2,1,1}}{z}+\bar{s}_{-2,1,1,1}+\bar{s}_{1,-2,1,1}-\frac{s_{-2,1,1}}{z}+s_{-2,1,1,1}+s_{
   1,-2,1,1}
\end{dmath}

%%%%%%%%%%%%%%%%%%%%%%%%%%%%%%%%%%%%%%%%%%%%%%%%%%%%%%%%%%%%%%%

\begin{dmath}
\bar{s}_1 s_{1,-2,1}+s_1 \bar{s}_{1,-2,1} =
   \frac{\bar{s}_{1,1}}{z^3}-\frac{-\bar{s}_{-3,1}+\bar{s}_{-2,1,1}-\bar{s}_{1,-2,1}-\frac{21}{8} \zeta
   _3 \bar{s}_{-1}+\frac{5}{8} \zeta _3 \bar{s}_1+\frac{1}{6} \pi ^2
   \bar{s}_{-2}}{z}-\bar{s}_{-4,1}+\bar{s}_{-3,1,1}-\bar{s}_{2,-2,1}+2 \bar{s}_{1,1,-2,1}-\frac{11}{8}
   \zeta _3 \bar{s}_{-2}-\frac{5}{8} \zeta _3 \bar{s}_2+\frac{\bar{s}_1}{z^4}+\frac{1}{6} \pi ^2
   \bar{s}_{-3}+\frac{1}{480} \pi ^4
   \bar{s}_1-\frac{s_{1,1}}{z^3}+\frac{-s_{-3,1}+s_{-2,1,1}-s_{1,-2,1}-\frac{21}{8} \zeta _3
   s_{-1}+\frac{5 \zeta _3 s_1}{8}+\frac{1}{6} \pi ^2 s_{-2}}{z}-s_{-4,1}+s_{-3,1,1}-s_{2,-2,1}+2
   s_{1,1,-2,1}-\frac{\pi ^2 \zeta _3}{6}-\frac{49 \zeta _5}{16}-\frac{11}{8} \zeta _3 s_{-2}-\frac{5
   \zeta _3 s_2}{8}+\frac{s_1}{z^4}+\frac{1}{6} \pi ^2 s_{-3}+\frac{\pi ^4 s_1}{480}+\frac{4 \zeta
   _3}{z^2}
\end{dmath}
and its the most complicated part is given by 
\begin{dmath}
\bar{s}_1 s_{1,-2,1}+s_1 \bar{s}_{1,-2,1}\to -\frac{\bar{s}_{-2,1,1}}{z}+\frac{\bar{s}_{1,-2,1}}{z}+2
   \bar{s}_{1,1,-2,1}+\frac{s_{-2,1,1}}{z}-\frac{s_{1,-2,1}}{z}+2 s_{1,1,-2,1}
\end{dmath}

%%%%%%%%%%%%%%%%%%%%%%%%%%%%%%%%%%%%%%%%%%%%%%%%%%%%%%%%%%%%

\begin{dmath}
\bar{s}_1 s_{1,1,-2}+s_1 \bar{s}_{1,1,-2}= -\frac{1}{2} \zeta _3
   \bar{s}_{1,1}-\frac{\bar{s}_{-2,1}}{z^2}+\frac{\bar{s}_{-3,1}}{z}+\frac{\pi ^2 \bar{s}_{1,-1}}{4
   z}-\frac{\pi ^2 \bar{s}_{1,1}}{12
   z}-\frac{\bar{s}_{1,-2,1}}{z}+\frac{\bar{s}_{1,1,-2}}{z}-\bar{s}_{-4,1}-\frac{1}{12} \pi ^2
   \bar{s}_{1,-2}-\frac{1}{12} \pi ^2 \bar{s}_{1,2}-\frac{1}{4} \pi ^2 \bar{s}_{2,-1}+\frac{1}{12} \pi ^2
   \bar{s}_{2,1}+\bar{s}_{1,-3,1}-\bar{s}_{1,2,-2}+\bar{s}_{2,-2,1}-\bar{s}_{2,1,-2}-\bar{s}_{1,1,-2,1}+3
   \bar{s}_{1,1,1,-2}-\frac{3}{2} \zeta _3 \bar{s}_{-2}+\frac{1}{2} \zeta _3
   \bar{s}_2+\frac{\bar{s}_1}{z^4}+\frac{\pi ^2 \bar{s}_{-1}}{4 z^2}-\frac{\pi ^2 \bar{s}_1}{12
   z^2}+\frac{7 \zeta _3 \bar{s}_{-1}}{4 z}-\frac{3 \zeta _3 \bar{s}_1}{4 z}-\frac{\pi ^2 \bar{s}_{-2}}{6
   z}-\frac{\pi ^2 \ln _2 \bar{s}_{-1}}{4 z}+\frac{\pi ^2 \ln _2 \bar{s}_1}{4 z}+\frac{1}{6} \pi ^2
   \bar{s}_{-3}-\frac{7}{240} \pi ^4 \bar{s}_1+\frac{1}{4} \pi ^2 \ln _2 \bar{s}_{-2}-\frac{1}{4} \pi ^2
   \ln _2 \bar{s}_2-\frac{1}{2} \zeta _3 s_{1,1}-\frac{s_{-2,1}}{z^2}-\frac{s_{-3,1}}{z}-\frac{\pi ^2
   s_{1,-1}}{4 z}+\frac{\pi ^2 s_{1,1}}{12
   z}+\frac{s_{1,-2,1}}{z}-\frac{s_{1,1,-2}}{z}-s_{-4,1}-\frac{1}{12} \pi ^2 s_{1,-2}-\frac{1}{12} \pi ^2
   s_{1,2}-\frac{1}{4} \pi ^2 s_{2,-1}+\frac{1}{12} \pi ^2
   s_{2,1}+s_{1,-3,1}-s_{1,2,-2}+s_{2,-2,1}-s_{2,1,-2}-s_{1,1,-2,1}+3 s_{1,1,1,-2}-\frac{3 \pi ^2 \zeta
   _3}{8}-\frac{39 \zeta _5}{16}-\frac{3}{2} \zeta _3 s_{-2}+\frac{\zeta _3
   s_2}{2}+\frac{s_1}{z^4}+\frac{\pi ^2 s_{-1}}{4 z^2}-\frac{\pi ^2 s_1}{12 z^2}-\frac{7 \zeta _3
   s_{-1}}{4 z}+\frac{3 \zeta _3 s_1}{4 z}+\frac{\pi ^2 s_{-2}}{6 z}+\frac{\pi ^2 s_{-1} \ln _2}{4
   z}-\frac{\pi ^2 s_1 \ln _2}{4 z}+\frac{1}{6} \pi ^2 s_{-3}-\frac{7 \pi ^4 s_1}{240}+\frac{1}{4} \pi ^2
   s_{-2} \ln _2-\frac{1}{4} \pi ^2 s_2 \ln _2+\frac{2 \zeta _3}{z^2}
\end{dmath}
and its the most complicated part is given by 
\begin{dmath}
\bar{s}_1 s_{1,1,-2}+s_1 \bar{s}_{1,1,-2}\to
   -\frac{\bar{s}_{1,-2,1}}{z}+\frac{\bar{s}_{1,1,-2}}{z}-\bar{s}_{1,1,-2,1}+3
   \bar{s}_{1,1,1,-2}+\frac{s_{1,-2,1}}{z}-\frac{s_{1,1,-2}}{z}-s_{1,1,-2,1}+3 s_{1,1,1,-2}
\end{dmath}

%%%%%%%%%%%%%%%%%%%%%%%%%%%%%%%%%%%%%%%%%%%%%%%%%%%%%%%%%%%%%%%
\begin{dmath}
s_{1,1} \bar{s}_{-2,1}+s_{-2,1} \bar{s}_{1,1}=
   \frac{\bar{s}_{1,1}}{z^3}+\frac{\bar{s}_{-2,1}}{z^2}-\frac{\bar{s}_{2,1}}{z^2}+\frac{2
   \bar{s}_{1,1,1}}{z^2}-\frac{\bar{s}_{-3,1}}{z}+\frac{\bar{s}_{-2,1,1}}{z}+\frac{\bar{s}_{1,-2,1}}{z}+\bar{s}_{-3,1,1}-\bar{s}_{2,-2,1}-\bar{s}_{-2,1,1,1}+\bar{s}_{1,1,-2,1}-\frac{3}{8} \zeta _3
   \bar{s}_{-2}-\frac{5}{8} \zeta _3 \bar{s}_2+\frac{\pi ^2 \bar{s}_1}{6 z^2}-\frac{21 \zeta _3
   \bar{s}_{-1}}{8 z}+\frac{5 \zeta _3 \bar{s}_1}{8 z}+\frac{\pi ^2 \bar{s}_{-2}}{6 z}+\frac{1}{480} \pi
   ^4 \bar{s}_1-\frac{s_{1,1}}{z^3}+\frac{s_{-2,1}}{z^2}-\frac{s_{2,1}}{z^2}+\frac{2
   s_{1,1,1}}{z^2}+\frac{s_{-3,1}}{z}-\frac{s_{-2,1,1}}{z}-\frac{s_{1,-2,1}}{z}+s_{-3,1,1}-s_{2,-2,1}-s_{
   -2,1,1,1}+s_{1,1,-2,1}-\frac{13 \zeta _5}{16}-\frac{3}{8} \zeta _3 s_{-2}-\frac{5 \zeta _3
   s_2}{8}+\frac{\pi ^2 s_1}{6 z^2}+\frac{21 \zeta _3 s_{-1}}{8 z}-\frac{5 \zeta _3 s_1}{8 z}-\frac{\pi
   ^2 s_{-2}}{6 z}+\frac{\pi ^4 s_1}{480}+\frac{2 \zeta _3}{z^2}
\end{dmath}
and its the most complicated part is given by 
\begin{dmath}
s_{1,1} \bar{s}_{-2,1}+s_{-2,1} \bar{s}_{1,1}\to
   \frac{\bar{s}_{-2,1,1}}{z}+\frac{\bar{s}_{1,-2,1}}{z}-\bar{s}_{-2,1,1,1}+\bar{s}_{1,1,-2,1}-\frac{s_{-
   2,1,1}}{z}-\frac{s_{1,-2,1}}{z}-s_{-2,1,1,1}+s_{1,1,-2,1}
\end{dmath}

%%%%%%%%%%%%%%%%%%%%%%%%%%%%%%%%%%%%%%%%%%%%%%%%%%%%%%%%%%%%%%%%

\begin{dmath}
s_{1,1} \bar{s}_{1,-2}+s_{1,-2} \bar{s}_{1,1}= -\zeta _3
   \bar{s}_{1,1}+\frac{\bar{s}_{1,1}}{z^3}+\frac{\bar{s}_{1,-2}}{z^2}+\frac{\bar{s}_{-3,1}}{z}-\frac{\bar
   {s}_{2,-2}}{z}-\frac{\bar{s}_{-2,1,1}}{z}-\frac{\bar{s}_{1,-2,1}}{z}+\frac{2
   \bar{s}_{1,1,-2}}{z}-\bar{s}_{-4,1}-\frac{1}{6} \pi ^2 \bar{s}_{1,-2}-\frac{1}{6} \pi ^2
   \bar{s}_{1,2}+\bar{s}_{3,-2}+\bar{s}_{-3,1,1}+2 \bar{s}_{1,-3,1}-2 \bar{s}_{1,2,-2}+\bar{s}_{2,-2,1}-2
   \bar{s}_{2,1,-2}-\bar{s}_{1,-2,1,1}-2 \bar{s}_{1,1,-2,1}+3 \bar{s}_{1,1,1,-2}-\frac{7}{8} \zeta _3
   \bar{s}_{-2}+\frac{3}{8} \zeta _3 \bar{s}_2+\frac{7 \zeta _3 \bar{s}_{-1}}{8 z}-\frac{3 \zeta _3
   \bar{s}_1}{8 z}-\frac{\pi ^2 \bar{s}_{-2}}{12 z}-\frac{\pi ^2 \bar{s}_2}{12 z}+\frac{1}{12} \pi ^2
   \bar{s}_{-3}-\frac{11}{288} \pi ^4 \bar{s}_1+\frac{1}{12} \pi ^2 \bar{s}_3-\zeta _3
   s_{1,1}-\frac{s_{1,1}}{z^3}+\frac{s_{1,-2}}{z^2}-\frac{s_{-3,1}}{z}+\frac{s_{2,-2}}{z}+\frac{s_{-2,1,1
   }}{z}+\frac{s_{1,-2,1}}{z}-\frac{2 s_{1,1,-2}}{z}-s_{-4,1}-\frac{1}{6} \pi ^2 s_{1,-2}-\frac{1}{6} \pi
   ^2 s_{1,2}+s_{3,-2}+s_{-3,1,1}+2 s_{1,-3,1}-2 s_{1,2,-2}+s_{2,-2,1}-2 s_{2,1,-2}-s_{1,-2,1,1}-2
   s_{1,1,-2,1}+3 s_{1,1,1,-2}-\frac{5 \pi ^2 \zeta _3}{12}-\frac{21 \zeta _5}{16}-\frac{7}{8} \zeta _3
   s_{-2}+\frac{3 \zeta _3 s_2}{8}-\frac{7 \zeta _3 s_{-1}}{8 z}+\frac{3 \zeta _3 s_1}{8 z}+\frac{\pi ^2
   s_{-2}}{12 z}+\frac{\pi ^2 s_2}{12 z}+\frac{1}{12} \pi ^2 s_{-3}-\frac{11 \pi ^4 s_1}{288}+\frac{\pi
   ^2 s_3}{12}+\frac{2 \zeta _3}{z^2}
\end{dmath}
and its the most complicated part is given by 
\begin{dmath}
s_{1,1} \bar{s}_{1,-2}+s_{1,-2} \bar{s}_{1,1}\to
   -\frac{\bar{s}_{-2,1,1}}{z}-\frac{\bar{s}_{1,-2,1}}{z}+\frac{2
   \bar{s}_{1,1,-2}}{z}-\bar{s}_{1,-2,1,1}-2 \bar{s}_{1,1,-2,1}+3
   \bar{s}_{1,1,1,-2}+\frac{s_{-2,1,1}}{z}+\frac{s_{1,-2,1}}{z}-\frac{2 s_{1,1,-2}}{z}-s_{1,-2,1,1}-2
   s_{1,1,-2,1}+3 s_{1,1,1,-2}
\end{dmath}

%%%%%%%%%%%%%%%%%%%%%%%%%%%%%%%%%%%%%%%%%%%%%%%%%%%%%%%%%%%%%%%%

\begin{dmath}
\bar{s}_{-2} s_{1,1,1}+s_{-2} \bar{s}_{1,1,1}= -\frac{1}{12} \pi ^2 s_{-3}-\frac{1}{4} \pi ^2 \ln _2
   s_{-2}-\frac{\pi ^2 s_{-2}}{12 z}-\frac{s_{-2}}{z^3}-\frac{\pi ^2 \ln _2 s_{-1}}{4 z}-\frac{\pi ^2
   s_{-1}}{4 z^2}+\frac{\pi ^2 \ln _2 s_1}{4 z}+\frac{\pi ^2 s_1}{12 z^2}-\frac{13 \pi ^4
   s_1}{1440}+\frac{1}{4} \pi ^2 \ln _2 s_2+\frac{\pi ^2 s_2}{12 z}+\frac{\pi ^2 s_3}{12}+\frac{5}{8}
   s_{-2} \zeta _3+\frac{21 s_{-1} \zeta _3}{8 z}-\frac{s_1 \zeta _3}{8 z}-\frac{s_2 \zeta _3}{8}-\frac{3
   \zeta _3}{z^2}-\frac{13 \pi ^2 \zeta _3}{48}+\frac{23 \zeta _5}{16}-\frac{1}{12} \pi ^2
   \bar{s}_{-3}-\frac{1}{4} \pi ^2 \ln _2 \bar{s}_{-2}+\frac{5}{8} \zeta _3 \bar{s}_{-2}+\frac{\pi ^2
   \bar{s}_{-2}}{12 z}+\frac{\bar{s}_{-2}}{z^3}+\frac{\pi ^2 \ln _2 \bar{s}_{-1}}{4 z}-\frac{21 \zeta _3
   \bar{s}_{-1}}{8 z}-\frac{\pi ^2 \bar{s}_{-1}}{4 z^2}-\frac{\pi ^2 \ln _2 \bar{s}_1}{4 z}+\frac{\zeta
   _3 \bar{s}_1}{8 z}+\frac{\pi ^2 \bar{s}_1}{12 z^2}-\frac{13 \pi ^4 \bar{s}_1}{1440}+\frac{1}{4} \pi ^2
   \ln _2 \bar{s}_2-\frac{1}{8} \zeta _3 \bar{s}_2-\frac{\pi ^2 \bar{s}_2}{12 z}+\frac{1}{12} \pi ^2
   \bar{s}_3+\frac{s_{1,-2}}{z^2}-\frac{1}{12} \pi ^2 s_{1,-2}+\frac{\pi ^2 s_{1,-1}}{4 z}-\frac{1}{2}
   \zeta _3 s_{1,1}-\frac{\pi ^2 s_{1,1}}{12 z}-\frac{1}{12} \pi ^2
   s_{1,2}+\frac{s_{2,-2}}{z}+\frac{1}{4} \pi ^2 s_{2,-1}-\frac{1}{12} \pi ^2
   s_{2,1}+s_{3,-2}+\frac{\bar{s}_{1,-2}}{z^2}-\frac{1}{12} \pi ^2 \bar{s}_{1,-2}-\frac{\pi ^2
   \bar{s}_{1,-1}}{4 z}-\frac{1}{2} \zeta _3 \bar{s}_{1,1}+\frac{\pi ^2 \bar{s}_{1,1}}{12 z}-\frac{1}{12}
   \pi ^2 \bar{s}_{1,2}-\frac{\bar{s}_{2,-2}}{z}+\frac{1}{4} \pi ^2 \bar{s}_{2,-1}-\frac{1}{12} \pi ^2
   \bar{s}_{2,1}+\bar{s}_{3,-2}+s_{-3,1,1}+s_{1,-3,1}-\frac{s_{1,1,-2}}{z}+\frac{s_{1,1,1}}{z^2}-s_{1,2,-
   2}-s_{2,1,-2}+\bar{s}_{-3,1,1}+\bar{s}_{1,-3,1}+\frac{\bar{s}_{1,1,-2}}{z}+\frac{\bar{s}_{1,1,1}}{z^2}
   -\bar{s}_{1,2,-2}-\bar{s}_{2,1,-2}-s_{-2,1,1,1}-s_{1,-2,1,1}-s_{1,1,-2,1}+s_{1,1,1,-2}-\bar{s}_{-2,1,1
   ,1}-\bar{s}_{1,-2,1,1}-\bar{s}_{1,1,-2,1}+\bar{s}_{1,1,1,-2}
\end{dmath}
and its the most complicated part is given by 
\begin{dmath}
\bar{s}_{-2} s_{1,1,1}+s_{-2} \bar{s}_{1,1,1}\to
   \frac{\bar{s}_{1,1,-2}}{z}-\bar{s}_{-2,1,1,1}-\bar{s}_{1,-2,1,1}-\bar{s}_{1,1,-2,1}+\bar{s}_{1,1,1,-2}
   -\frac{s_{1,1,-2}}{z}-s_{-2,1,1,1}-s_{1,-2,1,1}-s_{1,1,-2,1}+s_{1,1,1,-2}
\end{dmath}

\newpage
\section*{D. Ansatz for most complicated terms in NNLO eigenvalue}\label{Appn1}
 \renewcommand{\theequation}{D-\arabic{equation}}
  % redefine the command that creates the equation no.
  \setcounter{equation}{0}  % reset counter 

Based on the arguments presented in the main text we can write two types of the most complicated terms, the cross products 
\begin{eqnarray}
s_1 \bar{s}_{-2,1,1}, \;\;s_1 \bar{s}_{1,-2,1}, \;\;s_1 \bar{s}_{1,1,-2}, \;\;s_{1,1} \bar{s}_{-2,1}, \;\;s_{1,1} \bar{s}_{1,-2}, \;\;s_{-2} \bar{s}_{1,1,1} 
\end{eqnarray}
and the functions of one variable
\begin{eqnarray}
s_{1,1,1,-2}, \;\; s_{1,1,-2,1}, \;\; s_{1,-2,1,1}, \;\; s_{-2,1,1,1}
\end{eqnarray}
as well as   their counterpart with $z \leftrightarrow \bar{z}$. Here we  use a compact notation of $s_{\{\alpha\}}=S_{\{\alpha\}}(z)$ and   $\bar{s}_{\{\alpha\}}=S_{\{\alpha\}}(\bar{z})$. 
There are also simpler terms that include  harmonic sums of lower depth and  transcendental constants, which are to be treated separately. 
The corresponding ansatz discussed in eq.~(\ref{complexansatz}) read
\begin{eqnarray}\label{complexansatz2}
&&A_3(z,\bar{z})= a_1 ( s_{1} \bar{s}_{-2,1,1}+ \bar{s}_{1} s_{-2,1,1})+
a_2 ( s_{1}  \bar{s}_{1,-2,1} + \bar{s}_{1}  s_{1,-2,1}) \nonumber 
\\
&&
+a_3 ( s_{1} \bar{s}_{1,1,-2}+ \bar{s}_{1} s_{1,1,-2})
+a_4 (s_{1,1} \bar{s}_{-2,1}  +\bar{s}_{1,1} s_{-2,1} )
\nonumber 
\\
&&+a_5 ( s_{1,1}  \bar{s}_{1,-2}+ \bar{s}_{1,1}  s_{1,-2})
+a_6 ( s_{1,1,1}  \bar{s}_{-2}+\bar{s}_{1,1,1}  s_{-2})\nonumber 
\\
&& + \phi+\bar{\phi}, 
\end{eqnarray} 
where $\phi$ is given by 
\begin{eqnarray}
\phi= b_1 s_{1,1,1,-2}+b_2 s_{1,1,-2,1}+b_3 s_{1,-2,1,1}+b_4 s_{-2, 1,1,1}
\end{eqnarray}
and 
\begin{eqnarray}
\bar{\phi}= b_1 \bar{s}_{1,1,1,-2}+b_2 \bar{s}_{1,1,-2,1}+b_3 \bar{s}_{1,-2,1,1}+b_4 \bar{s}_{-2, 1,1,1}.
\end{eqnarray}

Firstly, we decompose $A_3(z,\bar{z})$ for $n=0$ using the reflection identities of Appendix B and compare this to the NNLO result of  Gromov, Levkovich-Maslyuk and Sizov~\cite{GROMOV}  for $n=0$ in eq.~(\ref{F3lin}), which has only one relevant most complicated  harmonic sum  $-256 \;\; s_{1, -2, 1, 1} -256 \;\; \bar{s}_{1, -2, 1, 1} $. All other harmonic sum $s_{ -2, 1, 1,1}$, $s_{ 1,1,-2, 1}$ and $s_{ 1,1,1,-2}$ and their $z \leftrightarrow \bar{z}$ counterparts are absent. This fixes four out of ten coefficients $a_i$ and $b_i$. Next, we perform pole decomposition of   $A_3(z,\bar{z})$ for $n=1$ using the shifted reflection identities in Appendix C and impose a condition the most complicated pure  harmonic sums are absent. This fixes another three coefficients and we are left with unknown $a_1$, $a_2$ and $a_4$, while the rest of the coefficients are expressed in terms of those three as follows 
\begin{eqnarray}
 a_3&\to &  -a_1+2 a_2+256 
 \\
 a_5&\to & a_1-a_2+a_4-128 \nonumber
 \\
 a_6 &\to & -a_1-2 a_4  \nonumber
 \\
 b_7 &\to & -384 + a_1 - 3 a_2 - a_4 \nonumber
 \\
 b_8 &\to & -2 a_2 - a_4 \nonumber
 \\
  b_9 &\to &  -128 - a_1 - a_2 - a_4 \nonumber
 \\
 b_{10}&\to &  -2 a_1 - a_4 \nonumber
\end{eqnarray}
This unambiguously fixes the rest of the terms for the pole decomposition for  $n=1$ resulting in 
\begin{eqnarray}\label{g2n1app}
g_2(z,-z)&=&\frac{128 s_{1,-2,1}}{z}-\frac{128 s_{-2,1,1}}{z}
+\frac{128 \bar{s}_{1,-2,1}}{(-z)}-\frac{128 \bar{s}_{-2,1,1}}{(-z)}
\nonumber 
\\
&& + \mathtt{simpler} \;\;\;  \mathtt{terms}
\end{eqnarray}

The remaining free coefficients can be fixed by implying an additional condition of the absence of $\ln^2 |z|$ and $\ln^3 |z|$ divergences as $|z| \to \infty$ in the ansatz $A_3(z,\bar{z})$ as was discussed in the main text. This sets $a_1=a_4=0$ and $a_2=-128$ and then 
\begin{eqnarray}
 a_3=a_5=a_6= b_7=b_9=b_{10}=0, \;\;\; b_8=256.
\end{eqnarray}
Plugging this into $A_3(z,\bar{z})$ in eq.~(\ref{complexansatz2}) we finally  get 
\begin{eqnarray} \label{ansatzSapp}
g_2(z,\bar{z})& =& -128 \left( s_{1} \bar{s}_{1,-2,1} + s_{1}  \bar{s}_{1,-2,1} \right)
+256 \left( s_{1,1,-2,1}+\bar{s}_{1,1,-2,1} \right) \nonumber 
\\
&& +  \mathtt{simpler} \;\;\;  \mathtt{terms}
\end{eqnarray}

\end{document}